\documentclass[twocolumn]{autart}
\usepackage[cmex10]{amsmath}
\usepackage{subfigure}
\usepackage{graphics}
\usepackage{epsfig}
\usepackage{subfig}
\usepackage{color}
\usepackage{amsfonts}
\usepackage{epstopdf}
\usepackage{amssymb}
\usepackage{amsmath}
\usepackage{verbatim}

\usepackage{newtxtext,newtxmath}
\usepackage{ntheorem}
\theoremstyle{plain}
\newtheorem{assm}{Assumption}

\newtheorem{lema}{Lemma}
\newtheorem{thme}{Theorem}
\newtheorem{remark}{Remark}
\newenvironment*{proof}{{\it Proof.}}{\hfill $\square$\par}

\newtheorem{define}{Definition}
\begin{document}

\begin{frontmatter}

\title{Safe Output Regulation  of Coupled Hyperbolic PDE-ODE Systems\thanksref{footnoteinfo}}
\thanks[footnoteinfo]{The material in this paper was not presented at any conference.}
\author[Xiamen]{Ji Wang}\ead{jiwang@xmu.edu.cn}~and
\author[SanDiego]{Miroslav Krstic}\ead{krstic@ucsd.edu}
\address[Xiamen]{Department of Automation, Xiamen University, Xiamen 361005, China}
\address[SanDiego]{Department of Mechanical and Aerospace Engineering, University of California, San Diego, La Jolla, CA 92093-0411, USA}


\begin{abstract}                          
This paper presents a safe output regulation control strategy for a class of systems modeled by a coupled
$2\times 2$ hyperbolic PDE-ODE structure, subject to fully distributed disturbances throughout the system. A state-feedback controller is developed by the {{nonovershooting backstepping}} method to simultaneously achieve output regulation and enforce safety constraints on the regulated output that is the state furthest from the control input. To handle unmeasurable states and external disturbances, an extended observer is designed. Explicit bounds on the estimation errors are derived and used to construct a robust safe regulator that accounts for the uncertainties.
The proposed control scheme guarantees that:
1) If the regulated output is initially within the safe region, it remains there; otherwise, it will be rescued to the safety region within a prescribed time;
2) The output tracking error converges to zero;
3) The observer accurately estimates both the distributed states and external disturbances, with estimation errors converging to zero exponentially;
4) All signals in the closed-loop system remain bounded.
The effectiveness of the proposed method is demonstrated through a UAV delivery scenario with a cable-suspended payload, where the payload is regulated to track a desired reference while avoiding collisions with barriers.
\end{abstract}

\end{frontmatter}

\section{Introduction}
\subsection{Output Regulation of Coupled Hyperbolic PDEs}
The output regulation problem seeks to design a control law that ensures the system output tracks desired references and/or rejects undesired disturbances. In recent years, considerable attention has been devoted to output regulation for infinite-dimensional systems governed by partial differential equations (PDEs). The Internal Model Principle (IMP), which provides the classical theoretical framework for robust output regulation, has been extensively investigated for infinite-dimensional systems \cite{Paunonen2010,Paunonen2014,Paunonen2016,Rebarber2003}. Unlike the conventional IMP-based output regulation design, the present work is developed within an observer-based backstepping framework. Specifically, output regulation theory is integrated with the backstepping methodology \cite{Krstic2008} to build the regulation controller, while an extended observer is constructed to estimate the unmeasured plant states and exogenous signals, where the observer can be designed independently, and its derived error bound is subsequently incorporated into the safe controller design.

Initial contributions toward backstepping-based output regulation for
$2\times 2$ hyperbolic PDEs appeared in \cite{Aamo2013}, upon the backstepping-based boundary stabilization of
$2\times 2$ hyperbolic PDEs presented in \cite{Vazquez2011Backstepping} and \cite{Coron2013Local}. Subsequent extensions to more general heterodirectional transport PDEs were proposed in \cite{Anfinsen2017}, \cite{Deutscher2017Output}, and \cite{2020Xu}. For wave-type PDEs, similar regulator designs were developed in \cite{2018Gabriel} and \cite{2018Gu}.
Research efforts were also directed at PDE-ODE interconnected systems, which naturally arise in many applications. Backstepping boundary control designs for interconnected
$2\times 2$ hyperbolic PDEs and ODEs were proposed in \cite{Meglio2017Stabilization}, and the ODE-PDE-ODE configurations were further addressed in \cite{J2017Control}, \cite{J2020delay}, \cite{Deutscher2018Output}, and \cite{Saba2019}. Subsequently, the output regulation of hyperbolic PDEs sandwiched between two ODEs was explored in \cite{Redaud}, and the regulation problems involving coupled wave PDE-ODE systems and hyperbolic PDEs coupled with nonlinear ODEs were investigated in \cite{Deutscher2021} and \cite{2023Irscheid}, respectively. {A more detailed literature review of backstepping output regulation/disturbance rejection in PDEs can be found in Sec. 3.5 of \cite{Vazquez2026Backstepping}.}
Despite progress in PDE output regulation, guaranteed safety remains unaddressed. Ensuring that outputs remain within safe operating regions is crucial for autonomous systems, thus the problem of safe output regulation of PDEs is essential and open.
\subsection{Safe Control}
Control Barrier Functions (CBFs) have emerged as a powerful tool for safe control design, as introduced in \cite{Ames2016}, where system states are constrained to remain within a designated safe region by ensuring the non-negativity of an appropriately constructed CBF. To enforce this constraint within the closed-loop system, CBF-based control strategies are typically combined with a quadratic program (QP) safety filter, which overrides potentially unsafe nominal control inputs to generate safe control actions.
High relative degree CBFs have been further developed in \cite{2016Nguyen}, \cite{2019Xiao}, and \cite{2021Xiao}, after the presence of non-overshooting control design techniques in \cite{KrsticBement2006}. Based on the tools developed in \cite{KrsticBement2006}, mean-square stabilization of stochastic nonlinear systems toward an equilibrium on the barrier was achieved in \cite{Li2020Mean}, and prescribed-time safety (PTSf)---which guarantees safety over a user-specified finite time horizon---was proposed in \cite{Abel2023}.

However, safe control of PDE systems remains largely underexplored. The first result in this area was presented in \cite{Koga2023Safe}, which proposed a CBF-based boundary control scheme for a Stefan model involving parabolic PDEs with actuator dynamics. Adaptive safe control for hyperbolic PDEs was proposed in \cite{supp}. Yet, these approaches address full-state control but do not address output regulation in the presence of unmeasured states and external disturbances.
\subsection{Main Contributions}
1) In contrast to traditional output regulation of hyperbolic PDEs \cite{Aamo2013}, \cite{Deutscher2017Output}, \cite{Deutscher2021}, \cite{Redaud}, our approach not only ensures that the system output tracks the desired trajectory but also guarantees that it stays within a prescribed safe region.\\
2) Compared to our previous work \cite{supp} on safe control of coupled hyperbolic PDEs, this paper presents several improvements:
a) Beyond the state-feedback framework, a safe output-feedback controller is developed, which accounts for unmeasured PDE states and external disturbances during the safe regulation;
b) Instead of restricting safety to positivity, a general barrier function $h$ is introduced, enabling broader and more flexible safety specifications;
c) {If the system starts outside the safe set, the proposed controller guarantees that it will be rescued to the safety within a prescribed time that can be selected as a design parameter by the user.}\\
3) To the best of our knowledge, this work presents the first result on safe output regulation for PDE systems. 
\subsection{Notation}
\begin{itemize}
\item The symbol $\mathbb N$ denotes the set $\{1,2,\cdots\}$.

\item {We use the notation $L^\infty(0, 1)$ for the standard Banach space of essentially bounded, measurable functions $f:(0,1) \to \mathbb{R}$, equipped with the norm $\|f\|_\infty:=\text{ess sup}_{x \in (0,1)} |f(x)| < +\infty$ for $f \in L^\infty(0, 1)$.}

\item  We use the notation $L^2(0, 1)$ for the standard space of the equivalence
class of square-integrable, measurable functions $f:(0,1)\to\mathbb R$, with $\|f\|:=\left(\int_0^1 f(x)^2 dx\right)^{\frac{1}{2}}<+\infty$ for $f \in L^2(0, 1)$.

\item  The notation
$\sup_{t\ge t_0}|f(t)|<\infty$
means that there exists a constant $M_h>0$ such that
$|f(t)|\le M_h,~\forall t\ge t_0 $.

\item Let $u: \mathbb R_+\times [0,1]\rightarrow \mathbb R$  be given. We use the notation $u[t]$ to denote the profile of $u$ at certain $t\ge 0$, i.e., $u[t]=u(x,t)$ for all $x\in[0,1]$.

\item Let $|Y|$ be Euclidean norm of the vector $Y$.

        \item We define $\underline x_j:= [x_1,x_2,\cdots,x_j]^T$.
\end{itemize}
For ease of presentation,  we omit or simplify the arguments of functions and functionals when no confusion arises. Besides, if $a>b$ happens in $\sum_{i=a}^{b}$ of this paper, it means that the result is zero.
\section{Problem Formulation}
\subsection{Plant}\label{sec:plant}
The considered plant  is
\begin{align}
\dot Y(t) &= AY(t) + Bw(0,t)+G_1d(t),\label{eq:o1}\\
{z_t}(x,t) &=  - q_1{z_x}(x,t)+{d_1}w(x,t)+G_2(x)d(t),\label{eq:o2}\\
{w_t}(x,t) &= q_2{w_x}(x,t)+{d_2}z(x,t)+G_3(x)d(t),\label{eq:o3}\\
z(0,t) &=pw(0,t)+CY(t)+G_4d(t),\label{eq:o4}\\
w(1,t) &=  qz(1,t)+G_5d(t)+ U(t),\label{eq:o5}
\end{align}
$\forall (x,t) \in [0,1]\times[0,\infty)$, where the function $U(t)$ is the control input  to be designed. The scalars $z(x,t)\in \mathbb{R}, w(x,t)\in \mathbb{R}$ are states of the PDEs, $Y^{T}(t)=[y_1,y_2,\cdots,y_n]\in \mathbb{R}^{n}$ are ODE states. The disturbances $d(t)\in \mathbb R^{m_d\times 1}$ are unmeasurable. The plant parameters $d_1,d_2$, $q$, the transport speed $q_1>0,q_2>0$, the matrices $G_1\in\mathbb R^{n\times m_d}, G_2, G_3\in (C[0,1])^{1\times m_d}$ regarding the disturbance input locations, and {$p\neq 0$}, are known and arbitrary.
The subsystem $Y$-ODE is strict-feedback linear, where the matrix $A$, the column vector $B$ are in the form of
\begin{align}
A=&\left(
  \begin{array}{cccccc}
    a_{1,1} & 1 & 0 & 0 & \cdots & 0 \\
    a_{2,1} & a_{2,2} & 1 & 0 & \cdots & 0 \\
     &  & \vdots &  &  &  \\
    a_{n-1,1} & a_{n-1,2} & a_{n-1,3} & a_{n-1,4} & \cdots & 1 \\
    a_{n,1} & a_{n,2}& a_{n,3} & a_{n,4} & \cdots & a_{n,n} \\
  \end{array}
\right),\label{eq:ABC}
\end{align}
and $B=(0,0,\cdots,b)^T$, with arbitrary constants $a_{i,j}$, and $b>0$ (without any loss of generality for $b<0$). This indicates that the $Y$-ODE is in the controllable form, which covers many practical models. {The matrix $C\in \mathbb R^{1\times n}$ is arbitrary. }

\subsection{Exogenous signal model}\label{sec:signal}
Like many articles on output regulation \cite{Deutscher2017Output}, the reference and disturbance signals are considered to be generated by a finite-dimensional exogenous model:
\begin{align}
\dot v(t) = S v(t),~~t>t_0,~~v(t_0)\in \mathbb R^{n_v}\label{eq:v}
\end{align}
where the spectrum $\sigma(S)$ of the known diagonalizable matrix $S\in\mathbb R^{n_v\times n_v}$ only contains eigenvalues on the
imaginary axis, which allows the modelling of bounded and persistently acting exogenous signals. In particular, this signal model can generate the exogenous
signals as constant or trigonometric functions of time as well
as linear combinations of both signal forms. Defining $S =
{\rm bdiag}(S_r, S_d)$ with $S_r\in \mathbb R^{n_r\times n_r}$, $S_d\in\mathbb R^{n_d\times n_d}$, and $v = {\rm col}(v_r, v_d)$ where $v_r\in\mathbb R^{n_r}$, $v_d\in\mathbb R^{n_d}$, one obtains the reference model
and the disturbance model, respectively,
\begin{align}
\dot v_r(t) = S_rv_r(t),~\dot v_d(t) = S_dv_d(t)\label{eq:vrvd}
\end{align}
where $n_r + n_d = n_v$. Also, we have
\begin{align}
&d(t)=P_dv(t)=\bar P_d v_d(t),~~P_d=\bar P_dE_d,\label{eq:vd}\\
&r(t)=P_rv(t)=\bar P_r v_r(t),~~P_r=\bar P_rE_r,\label{eq:r}
\end{align}
where $P_d\in\mathbb R^{m_d\times n_v}$, $P_r\in\mathbb R^{1\times n_v}$, $\bar P_d\in\mathbb R^{m_d\times n_d}$, $\bar P_r\in\mathbb R^{1\times n_r}$, $E_r=[I_r,0]\in \mathbb R^{n_r\times n_v}$, $E_d=[0,I_d]\in \mathbb R^{n_d\times n_v}$ where $I_d$, $I_r$ are identity matrices with dimensions of $n_d$, $n_r$, respectively.
\begin{assm}\label{as:obex}
{The pairs $(S_d,\bar P_d)$ and $(S_r,\bar P_r)$ are observable.}
\end{assm}
Please note that, since the exogenous signal model is neutrally stable, detectability and observability are equivalent.
\subsection{Control Objective}\label{sec:con}
{There are two measurable outputs in the considered plant: one is $z(1,t)$ that will be utilized in the extended observer for both unmeasurable states and exogenous signals, and the other one is $y_1=C_1 Y(t)$ that is the output to be safely regulated, which is described below.} The vector $C_1=[1,0,\cdots,0]$ satisfies the following assumption.
\begin{assm}\label{as:ob}
{The pair $(A,C_1)$ is observable. }
\end{assm}
{The control objective is to ensure that the output tracking error $e(t)=y_1(t)-r(t)$ converges to zero while staying in the time-varying safe region in the sense of $
e(t) \in \mathcal{C}(t), \forall t \ge \bar t_0$, where $\mathcal{C}(t)$ is a time-dependent safe set
$
\mathcal{C}(t) := \{ e \in \mathbb{R} \mid h(e,t) \ge 0 \}
$
defined by
the time-varying barrier function $h(e(t),t)$, and where the initial regulation time
\begin{align}
    \bar t_0=t_0+\frac{1}{q_2}\label{eq:bt0}
\end{align}
is introduced due to the delay property of the transport PDE. 
Specifically, there is no control influence on the distal ODE over the time interval $[t_0,\bar t_0]$, where $t_0$ is the initial time.
We impose the following assumption regarding the time-varying barrier function $h$ for the output signal.
\begin{assm}\label{as:h}
The time-varying barrier function $h$
is $n$ times continuously differentiable with respect to $e,t$.
Moreover, all its mixed partial derivatives of positive order up
to $n$ are uniformly bounded, i.e.,
$
\sup_{\substack{e\in\mathbb{R},\,t\geq t_0\\
j,k\in\{0,\ldots,n\},\,1\leq j+k\leq n}}
\left|
\frac{\partial^{j+k}h(e,t)}
{\partial e^j\partial t^k}
\right|<\infty.
$
There also exists a constant $\underline{\vartheta}>0$ such that
$\left|\frac{\partial h(e,t)}{\partial e}\right|
\geq\underline{\vartheta},
\qquad
\forall e\in\mathbb{R},\quad t\geq t_0.$
Besides,
$\sup_{t\geq t_0}|h(e(t),t)|<\infty
\Longrightarrow
\sup_{t\geq t_0}|e(t)|<\infty,$
and
$\lim_{t\rightarrow\infty}h(e(t),t)=0
\Longrightarrow
\lim_{t\rightarrow\infty}e(t)=0$.
\end{assm}
{Focusing on 
$y_1$ is only for ease of exposition and does not restrict the applicability of the result. The result is not restricted to systems whose first state only is constrained. The result is applicable to all systems diffeomorphically transformable into the form \eqref{eq:o1}, \eqref{eq:ABC}, which is equivalent to the input-output feedback linearization-like treatment with high-order CBFs (HOCBFs). And the result is also applicable, with an observer, to all systems transformable into \eqref{eq:o1} with a transformation acting not only on the measured state but on the entire (unmeasured) state. An example is given in the following remark.
} }
{\begin{remark} [Applicability]
  \rm {The control design is applicable to the case where the distal ODE subsystems (denoted as the $X$--ODE) are described by general matrices $\bar A\in\mathbb R^n,\bar B\in\mathbb R^{n\times 1},\bar C\in\mathbb R^{1\times n}$, assuming that the pair $(\bar A,\bar C)$ is observable and that the $X$--ODE has relative degree $n$. Such systems are also covered by the framework considered in this paper. 
Indeed, the $X$--ODE can be transformed into the $Y$--ODE studied here via the coordinate transformation
$Y(t) = R X(t)$, $R = [\bar C^T, (\bar C\bar A)^T,\ldots, (\bar C\bar A^{n-1})^T]^T$.
Since $(\bar A,\bar C)$ is observable, the matrix $R$ is nonsingular. The relative degree $n$ that implies that $\bar C\bar A^i\bar B=0$ for $i=0,\cdots,n-2$ and $\bar C\bar A^{n-1}\bar B\neq0$. Then it is ensured that the transformed system's system matrix, which is in the companion (observable canonical) form whose last row consists of the coefficients of the characteristic polynomial of $\bar A$ according to Hamilton-Cayley theorem, is covered by the structure $A$ in \eqref{eq:ABC}, and the resulting input and output matrices also match $B,C_1$ matrices considered in this paper. 
Furthermore, the system output is preserved under the transformation, namely,
$\bar CX = C_1 Y=y_1$.
Therefore, the safe regulation of the $Y$--ODE output achieved by the following control design directly implies the safe regulation of the original $X$--ODE output.}
\end{remark}}
{In the derivations below, we work under the standing strong assumption that the plant, and later the corresponding closed-loop and observer systems, admit classical solutions on the time interval under consideration, so that all temporal and spatial derivatives appearing in the PDEs and all boundary traces used in the control law and proofs are well defined pointwise.}
\section{Nominal Safe Control Design}\label{sec:TandB}
Applying \eqref{eq:vd},  then \eqref{eq:o1}--\eqref{eq:o5} are rewritten as
\begin{align}
\dot Y(t) &= AY(t) + Bw(0,t)+\acute G_1v(t),\label{eq:o1a}\\
{z_t}(x,t) &=  - q_1{z_x}(x,t)+{d_1}w(x,t)+\acute G_2(x)v(t),\label{eq:o2a}\\
{w_t}(x,t) &= q_2{w_x}(x,t)+{d_2}z(x,t)+\acute G_3(x)v(t),\label{eq:o3a}\\
z(0,t) &=pw(0,t)+CY(t)+\acute G_4v(t),\label{eq:o4a}\\
w(1,t) &=  qz(1,t)+\acute G_5v(t)+ U(t),\label{eq:o5a}
\end{align}
where $\acute G_i=G_iP_d,~i=1,\cdots,5$. The matrices $\acute G_i\in\mathbb R^{1\times n_v}, i=2,\cdots,5$ and $\acute G_1:=[\acute g_1;\cdots;\acute g_n]\in \mathbb R^{n\times n_v}$ with $\acute g_j\in\mathbb R^{1\times n_v}, j=1,\cdots,n$, are known.
The control design consists of three transformations, each serving a distinct purpose. The first transformation converts the output regulation problem into a tracking-error system with a chain-of-integrators structure, which facilitates the subsequent barrier-function design. The second transformation introduces the high-order barrier functions for safety enforcement. The third transformation is applied to remove the PDE coupling and exogenous-signal dependence, thereby transforming the original system into a target system for which the safe controller can then be constructed.
\subsection{First transformation}\label{sec:tran1}
First, we propose the following transformation to derive and rewrite the tracking error dynamics in a form that will be helpful for the next transformation to barrier functions,
{
\begin{align}
Z(t)=T_zY(t)+T_vv(t),\label{eq:Tran-z}
\end{align}
where $Z(t)=[z_1,\cdots,z_n]^T$ and the matrix $T_z\in\mathbb R^{n\times n}$ is
\begin{align}\label{eq:Tz}
T_z=\left(
      \begin{array}{ccccc}
        1 & 0 & 0 & 0 &0\\
        \varrho_{1,1} & 1 & 0 & 0 &0\\
       \varrho_{2,1} & \varrho_{2,2} & 1 & 0&0 \\
       \vdots\\
         \varrho_{n-1,1} & \varrho_{n-1,2} & \cdots & \varrho_{n-1,n-1} & 1 \\
      \end{array}
    \right)
\end{align}
with the constants $\varrho_{i,j}$ defined by
\begin{align}
&\varrho_{1,1}=a_{1,1},\label{eq:vqrrho11}\\
&\varrho_{2,1}= a_{2,1}+\varrho_{1,1}a_{1,1},~~\varrho_{2,2}=a_{2,2}+\varrho_{1,1}\label{eq:vqrrho21}
\end{align}
and, for $i=3,\cdots,n$, by
\begin{align}
  &\varrho_{i,1}= a_{i,1} +\sum_{j=1}^{i-1}\varrho_{i-1,j}a_{j,1},\label{eq:varrhoi1} \\
 &\varrho_{i,\imath}=a_{i,\imath}+\varrho_{i-1,\imath-1}+\sum_{j=\imath}^{i-1}\varrho_{i-1,j}a_{j,\imath},~\forall\imath=2,\cdots, i-1,\\
 &\varrho_{i,i}=a_{i,i}+\varrho_{i-1,i-1},\label{eq:varrho11}
\end{align}
and where the matrix $T_v\in\mathbb R^{n\times n_v}$ is
\begin{align}
T_v=-[P_r; \lambda_1+P_rS^{1};\cdots;\lambda_{n-1}+P_rS^{n-1}]\label{eq:Tv}
\end{align}
with the vectors $\lambda_{i}$ defined by
\begin{align}
&\lambda_0=0,\label{eq:lam0}\\
 &\lambda_i=-\sum_{j=1}^{i-1}\varrho_{i-1,j}\acute g_j -\acute g_i +\lambda_{i-1}S~~\forall~i=1,\cdots, n.\label{eq:lami}
\end{align}
Applying the transformation \eqref{eq:Tran-z}, we now convert the $Y$-ODE \eqref{eq:o1a} into
\begin{align}
\dot Z(t)=A_{\rm z} Z(t)+Bw(0,t)+BK^TY(t)-\bar G_0 v(t)\label{eq:ZA}
\end{align}
{where $T_zB=B$ is recalled}, and 
\begin{align}
A_{\rm z}=\left(
  \begin{array}{ccccccc}
    0 & 1 & 0 & 0 & \cdots & 0& 0 \\
    0 & 0 & 1 & 0 & \cdots & 0& 0 \\
     0 & 0 & 0 & 1 & \cdots & 0& 0 \\
     &  & \vdots &  &  &  &\\
    0 & 0 & 0 & 0 & \cdots & 0 &1 \\
    0 & 0 & 0 & \cdots & 0  &0 &0 \\
  \end{array}
\right)\label{eq:Az}
\end{align}
and where
\begin{align}
K^T=\frac{1}{b}[\varrho_{n,1},\cdots,\varrho_{n,n}]_{1\times n}, \label{eq:K}
\end{align}
with $\bar G_0=\frac{1}{b}B(\lambda_n+P_rS^{n})$.}
Regarding $Z$-ODE in \eqref{eq:ZA},  $z_1$ is the tracking error that is to be regulated to zero, i.e.,
 \begin{align}
 z_1(t)=e(t)\label{eq:z1e}
 \end{align}
and the system matrix $A_{\rm z}$ represents a chain-of-integrators. The details of the first transformation are shown in Appendix \ref{sec:trans1Ap}.
\subsection{Second transformation}\label{sec:tran2}
For the purpose of establishing barrier functions for the distal ODE, inspired by \cite{Abel2023},  we propose the following transformation
\begin{align}
&h_1(z_1(t),t)=h(e(t),t)+\sigma(t),\label{eq:h1}\\
&h_i(\underline z_i,t)=\sum_{\jmath=1}^{i-1} \frac{\partial h_{i-1}}{\partial z_{\jmath}}z_{\jmath+1}+\frac{\partial h_{i-1}}{\partial t}+k_{i-1}h_{i-1},\label{eq:hi}
\end{align}
for $i=2,3,\cdots,n$, with
          \begin{align}\sigma(t)= \begin{cases}
\begin{cases}
e^{\frac{1}{t_a^2}}({-h(e(\bar t_0),\bar t_0)+\epsilon})e^{\frac{-1}{(t-\bar t_0-t_a)^2}}, t\in[t_0, \bar t_0+t_a), \\
0,~~~~~~~~~~~~~~~~~~~~~~~~~~~~~~~~~~~~~~ t\ge \bar t_0+t_a
\end{cases} \\  if~~~h(e(\bar t_0),\bar t_0)\le 0.  \\
0,~~ t\ge t_0,~~~~~ if~~ h(e(\bar t_0),\bar t_0)>0
\end{cases}\label{eq:sigma}
          \end{align}
where $t_a$ and $\epsilon$ are arbitrarily positive design parameters. The function $\sigma(t)$ is designed to address scenarios where the states are in the unsafe region at the initial regulation time $\bar t_0$. Specifically, when the initial state $e(\bar t_0)$ falls outside the original safe region, indicated by the condition $h(e(\bar{t}_0), \bar{t}_0) \leq 0$, a new barrier function is created to guide the state back to the original safe region. The constant $\epsilon$ indicates the distance from the safe barrier of the initial value of the new barrier function $h_1$, and
 $t_a$ is the upper bound of the time taken to rescue to the safe region. We only ensure that the state, which falls into the unsafe region at the initial regulation time $\bar t_0$, rescues to safety within a prescribed time $t_a$, but not prescribed-time regulation to the safe boundary. {Please note that even though $t_a$ can be assigned arbitrarily, a smaller value of $t_a$ leads to larger control gains, and thus a proper trade-off should be considered in practical implementation.} The evolution of $e(t)$ on $t\in[t_0,\bar t_0]$ can be expressed as the initial states as
\begin{align}
e(t_0+a)=C_1Z(t_0+a)\label{eq:icG}
\end{align}
$\forall a\in[0,\frac{1}{q_2}]$, where $Z(t_0+a)$, which is determined by the initial states $Y(t_0), w[t_0], z[t_0]$ and $v(t_0)$, denotes the predicted state of the $Z$-ODE, and its explicit expression is given by \eqref{eq:Zfuture} in Appendix \ref{sec:Yfuture}.
Inserting \eqref{eq:icG} into the barrier function $h(e,t)$, we then obtain the prediction mapping $\mathcal P$ 
\begin{align}
h(e(t_0+a),t_0+a)=\mathcal P(Y(t_0),z[t_0],w[t_0],v(t_0),a)\label{eq:mathp}
\end{align}
$\forall a\in[0,\frac{1}{q_2}]$.  Setting $a=\frac{1}{q_2}$, $h(e(\bar t_0),\bar t_0)$  in \eqref{eq:sigma} is then obtained as $h(e(\bar t_0),\bar t_0)=\mathcal P(Y(t_0),z[t_0],w[t_0],v(t_0),\frac{1}{q_2})$, depending on the initial states of the plant.

Note that $\sigma(t)$ is continuous and has continuous derivatives of all orders. Therefore, together with Assumption \ref{as:h}, $h_1(z_1(t),t)$ is $n$ times continuously differentiable with respect to each of its arguments, i.e., $z_1$ and $t$. Please also note that $\dot h(e(t),t)$ denotes the full derivative with respect to $t$ whose calculation uses the chain rule, and $\frac{\partial h(e(t),t)}{\partial t}$ denotes the partial derivative with respect to one of its variables $t$.
Define
\begin{align}
\frac{\partial h(e(t),t)}{\partial e(t)}:=\vartheta(e(t),t)=\vartheta(z_1(t),t),\label{eq:hy1}
\end{align}
because of $z_1(t)=e(t)$ \eqref{eq:z1e}. Considering the fact
\begin{align}
    \frac{\partial h_{n}}{\partial z_{n}}=\frac{\partial h_{n-1}}{\partial z_{n-1}}=\frac{\partial h_{n-2}}{\partial z_{n-2}}=\cdots=\frac{\partial h_{1}}{\partial z_{1}}=\frac{\partial h}{\partial e}=\vartheta\label{eq:hd}
\end{align}
that is obtained from \eqref{eq:hi}, using \eqref{eq:hi} for $i=n$, we have
$\dot h_n(\underline z_n,t)+k_nh_n
=b(\vartheta w(0,t)+f(\underline z_n(t),t)-\vartheta\frac{1}{b}(\lambda_n+P_rS^{n}) v(t)+\vartheta K^TY(t))$,
where
\begin{align}
&f(\underline z_n(t),t)=\frac{1}{b}\bigg(\sum_{\jmath=1}^{n-1} \frac{\partial h_{n}}{\partial z_{\jmath}}z_{\jmath+1}+ \frac{\partial h_{n}}{\partial t}\notag\\&+k_n[\sum_{\jmath=1}^{n-1} \frac{\partial h_{n-1}}{\partial z_{\jmath}}z_{\jmath+1}+ \frac{\partial h_{n-1}}{\partial t}+k_{n-1}h_{n-1}]\bigg).\label{eq:f}
\end{align}
Applying \eqref{eq:h1}, \eqref{eq:hi} to convert \eqref{eq:ZA} into
\begin{align}
\dot H(t)=&A_{\rm h} H(t)+B\bigg(f(\underline z_n(t),t)+\vartheta w(0,t)\notag\\&-\vartheta \frac{1}{b}(\lambda_n+P_rS^{n}) v(t)+\vartheta K^TY(t)\bigg)\label{eq:hA}
\end{align}
where {$H=[h_1,\cdots,h_n]^T$}, and where
\begin{align}
A_{\rm h}=\left(
  \begin{array}{ccccccc}
    -k_1 & 1 & 0 & 0 & \cdots & 0& 0 \\
    0 & -k_2 & 1 & 0 & \cdots & 0& 0 \\
     0 & 0 & -k_3 & 1 & \cdots & 0& 0 \\
     &  & \vdots &  &  &  &\\
    0 & 0 & 0 & 0 & \cdots & 0 &1 \\
    0 & 0 & 0 & \cdots & 0  &0 &-k_n \\
  \end{array}
\right).\label{eq:Ah}
\end{align}
{This chain structure $\dot H(t)=A_{\rm h} H(t)$, also known as the form of high relative-degree CBFs \cite{2016Nguyen}, \cite{2019Xiao}, was proposed in \cite{KrsticBement2006} for the nonovershooting control.}
\subsection{Third transformation}
In order to remove the in-domain coupling  destabilizing terms from  the $2\times 2$ hyperbolic PDE system, the extra terms in \eqref{eq:ZA}, and  {the dependence on the signal model state $v(t)$}, we propose the following backstepping transformations:
\begin{align}
&\alpha (x,t) =z(x,t) - \int_0^x {\phi}(x,y)z(y,t)dy\notag\\& -  \int_0^x {\varphi}(x,y)w(y,t)dy
-\gamma(x)Y(t)-\bar\gamma(x)v(t)- \bar\varsigma(x,t),\label{eq:contran1a}\\
&\beta (x,t) =w(x,t) -\int_0^x {\Psi}(x,y)z(y,t)dy\notag\\&  -\int_0^x {\Phi}(x,y)w(y,t)dy
-\lambda(x)Y(t)-\bar\lambda(x)v(t)-{\varsigma}(x,t)\label{eq:contran1b}
\end{align}
where $\phi,\varphi,\Psi,\Phi$ are the PDE backstepping kernels,
and $\gamma,\lambda$ are introduced to compensate the ODE coupling.  These kernels satisfy the equations given in  Appendix \ref{sec:ker}, which constitute a
well-known class of heterodirectional linear hyperbolic PDE--ODE systems,
whose well-posedness has been established in
\cite{Meglio2017Stabilization}. The functions $\bar\gamma,\bar\lambda$, which are the regulator kernels associated with the
exogenous signals, satisfy the regulator equations given by \eqref{eq:barlam1}--\eqref{eq:kerz} that constitute two independent linear matrix
initial-value problems with respect to the spatial variable $x$, whose explicit solution is given by \eqref{eq:barlambdasol}, \eqref{eq:bargammasol}.
Besides, the functions ${\varsigma}(x,t)$ and $\bar\varsigma(x,t)$ satisfy
\begin{align}
&\varsigma_t(x,t)=q_2{\varsigma}_x(x,t),~~\varsigma(0,t)=\frac{-1}{\vartheta(z_1(t),t) }f(\underline z_n,t),\label{eq:pt}\\
&\bar\varsigma_t(x,t)=-q_1\bar\varsigma_x(x,t),~~\bar\varsigma(0,t)=p*\varsigma(0,t).\label{eq:q0}
\end{align}
The solution $\varsigma(x,t)$ is
\begin{align}
\varsigma(x,t)=\frac{-1}{\vartheta(z_1(t+\frac{x}{q_2}),t+\frac{x}{q_2})}f\left(\underline z_n(t+\frac{x}{q_2}),t+\frac{x}{q_2}\right)\label{eq:pxt}\notag\\
\end{align}
where $\vartheta(z_1(t+\frac{x}{q_2}),t+\frac{x}{q_2})$ is nonzero according to Assumption \ref{as:h}. The solution \eqref{eq:pxt} is determined by the prediction $Z(t+\frac{x}{q_2})$ given by \eqref{eq:Zfuture} that is expressed by the current states $Y(t)$, $w[t]$, $z[t]$ and $v(t)$ recalling \eqref{eq:Tran-z} to replace $Z(t)$ with $Y(t)$.
By \eqref{eq:contran1a}, \eqref{eq:contran1b}, the system \eqref{eq:o2}--\eqref{eq:o5} with \eqref{eq:ZA} is converted into
\begin{align}
\dot H (t) =& A_{\rm h}H (t) + B\vartheta(z_1(t),t) \beta(0,t), \label{eq:targ5}\\
{\alpha _t}(x,t) =&  - {q_1}{\alpha _x}(x,t),~~
{\beta _t}(x,t) = {q_2}{\beta _x}(x,t),\label{eq:targ4}\\
  \alpha (0,t) =& p\beta  (0,t),~~ \beta(1,t) =0\label{eq:targ8}
\end{align}
by choosing the control input as
\begin{align}
U(t)=&-qz(1,t)+\int_0^1 {\Psi}(1,y)z(y,t)dy \notag\\&+\int_0^1 {\Phi}(1,y)w(y,t)dy+\lambda(1)Y(t)\notag\\&- (\acute G_5-\bar\lambda(1)) {v}(t)+\varsigma(1,t)\label{eq:U1}
\end{align}
where $\varsigma(1,t)$
is given by \eqref{eq:h1}--\eqref{eq:sigma}, \eqref{eq:f}, \eqref{eq:pxt}, \eqref{eq:Zfuture}, ${\Psi}(1,y), {\Phi}(1,y)$, $\lambda(1)$ are determined by \eqref{eq:traker1}--\eqref{eq:lamx}, and $\bar\lambda(1)$ is given by \eqref{eq:barlam1}, \eqref{eq:barlam2}, that is,
\begin{align}
\varsigma(1,t)=&\frac{-f(\wp_1(Z(t),z[t],w[t],v(t),\frac{1}{q_2}),t+\frac{1}{q_2})}{\vartheta(C_1\wp_1(Z(t),z[t],w[t],v(t),\frac{1}{q_2}),t+\frac{1}{q_2})}\notag\\
:=&g(Z(t),z[t],w[t],v(t),t)\notag\\=&g(T_zY(t)+T_vv(t),z[t],w[t],v(t),t)\label{eq:p1t}
\end{align}
where $g(\cdot)$ is introduced solely for notational convenience as a compact representation of the above expression, making the dependence of $\varsigma(1,t)$ on its arguments explicit.
\begin{remark}
   { \rm If actuator dynamics are included, the proposed safe output regulation scheme can be extended to an ODE–PDE–ODE configuration using the approach in \cite{supp}. It can also accommodate uncertain input delays in hyperbolic PDE–ODE cascades by integrating the method in \cite{2024delay}.}
\end{remark}
\subsection{Selection of safe design parameters}\label{sec:parameter}
The design parameters $k_{i}$, $i=1,2,\cdots,n-1$ satisfy
\begin{align}
k_{i}>\max\{0,\acute k_i(Y(t_0),z[t_0],w[t_0],v(t_0))\}\label{eq:kappai}
\end{align}
where
$\acute k_i=\frac{-1}{h_{i}(\underline z_{i}(\bar t_0),\bar t_0)}\big(\sum_{\jmath=1}^{i} \frac{\partial h_{i}}{\partial z_{\jmath}}z_{\jmath+1}(\bar t_0)+ \frac{\partial h_{i}}{\partial t}\big)$.
By \eqref{eq:Zfuture} and \eqref{eq:Tran-z},
$Z(\bar t_0)$ on the right-hand side can be expressed as the initial states $z[t_0],w[t_0],Y(t_0)$ and $v(t_0)$.
We write $\acute k_i(Y(t_0),z[t_0],w[t_0],v(t_0))$  to emphasize that it depends on the initial data of the overall plant. {The gain condition \eqref{eq:kappai} for the nonovershooting control} ensures the following lemma.
\begin{lema}\label{cl:zi0}
With the design parameters $k_i$, $i=1,\cdots,n-1$ satisfying \eqref{eq:kappai}, the high-relative-degree ODE CBFs are initialized positively, i.e., $h_i(\underline {z}_i(\bar t_0),\bar t_0)>0$, $i=1,\cdots,n$.
\end{lema}
\begin{proof}
According to \eqref{eq:h1} and \eqref{eq:sigma}, we know
$h_1(z_1(\bar t_0),\bar t_0)>0$.
Recalling \eqref{eq:hi}, we have
$h_i(\underline z_i(\bar t_0),\bar t_0)=\sum_{\jmath=1}^{i-1} \frac{\partial h_{i-1}}{\partial z_{\jmath}}z_{\jmath+1}(\bar t_0)+\frac{\partial h_{i-1}}{\partial t}+k_{i-1}h_{i-1}(\underline z_{i-1}(\bar t_0),\bar t_0)$, $i=2,3,\cdots,n$.
The lemma is thus obtained by recalling \eqref{eq:kappai}.
\end{proof}
{We give the following definition about the safe initial condition, which is the sufficient and necessary condition under which the output state stays in the safe region before the control input arrives at the distal ODE subsystem. It is the counterpart to the initial state restriction, i.e., Assumption 1, in the work \cite{Abel2024} about safe control with delays.}
\begin{define}\label{def:1}
{\rm The safe initial condition is defined as $h(e(t_0+a),t_0+a)=\mathcal P(Y(t_0),z[t_0],w[t_0],v(t_0),a)\ge 0$ for $a\in[0,\frac{1}{q_2}]$ and $h(e(\bar t_0),\bar t_0)=\mathcal P(Y(t_0),z[t_0],w[t_0],v(t_0),\frac{1}{q_2})\neq 0$, where the function $\mathcal P$ is the prediction mapping given by \eqref{eq:icG}, \eqref{eq:mathp}.}
\end{define}The condition $h(e(\bar t_0),\bar t_0)\neq 0$ means that the state does not stay at the safe boundary at the initial regulation time $t=\bar t_0$ for the distal ODE.

\subsection{Result of the state-feedback safe control}
\begin{thme}\label{th:th1}
{Assume that the closed-loop system consisting of \eqref{eq:o1}--\eqref{eq:o5} and \eqref{eq:U1} admits a classical solution.} Under Assumption \ref{as:h}, for initial data $w[t_0]\in L^\infty(0,1)$, $z[t_0]\in L^\infty(0,1)$, $Y(t_0)\in \mathbb R^n$, and for design parameters $k_i$, $i=1,\ldots,n-1$, satisfying \eqref{eq:kappai}, $k_n>0$, the following properties hold:\\
1) Tracking error $e(t)=y_1(t)-r(t)$ is convergent to zero and all states are bounded;\\
2) Safety is ensured in the sense that:\\
a) When the safe initial condition in Definition \ref{def:1} is satisfied, the safety is ensured from $t=t_0$, i.e., $h(e(t),t)\ge 0,~\forall t\ge t_0$;\\
b) When the safe initial condition in Definition \ref{def:1} is not satisfied, and $h(e(\bar t_0),\bar t_0)>0$, the safety is ensured from $t=\bar t_0$, i.e., $h(e(t),t)\ge 0,~\forall t\ge \bar t_0$;\\
c) When the safe initial condition in Definition \ref{def:1} is not satisfied, and $h(e(\bar t_0),\bar t_0)\le0$, the state will return and stay in the safe region no later than a finite time $\bar t_0+t_a$ where a constant $t_a>0$ can be arbitrarily assigned by users, i.e., $h(e(t),t)\ge 0,~\forall t\ge \bar t_0+t_a$.
\end{thme}
\begin{proof} 1) Recalling the target system \eqref{eq:targ5}--\eqref{eq:targ8} and Assumption \ref{as:h} indicates that $\vartheta$ is bounded, by the method of characteristics, it is obtained that $\beta[t],\alpha[t],H(t)$ are bounded all the time, and moreover, $\beta[t]\equiv 0$, $\alpha[t]\equiv 0$ for $t\ge t_0+\frac{1}{q_1}+\frac{1}{q_2}$ and $|H(t)|$ is exponentially convergent to zero recalling the fact that $A_{\rm h}$ defined in \eqref{eq:Ah} is Hurwitz. 

We then obtain from \eqref{eq:h1}, \eqref{eq:sigma}, and
Assumption~\ref{as:h} that $z_1(t)=e(t)$ converges to zero
and is therefore bounded. We next prove the boundedness of
$z_2,\ldots,z_n$ recursively. From \eqref{eq:hi} and
\eqref{eq:hd}, for $i=2,\ldots,n$, we have
\[
z_i
=
\frac{1}{\vartheta}
\left(
h_i
-\sum_{j=1}^{i-2}
\frac{\partial h_{i-1}}{\partial z_j}z_{j+1}
-\frac{\partial h_{i-1}}{\partial t}
-k_{i-1}h_{i-1}
\right),
\]
where the summation is understood to be zero for $i=2$.
For $i=2$, this relation gives
$z_2
=
\frac{1}{\vartheta}
\left(
h_2-\frac{\partial h_1}{\partial t}-k_1h_1
\right).$
The boundedness of $|H(t)|$ implies that $h_1$ and $h_2$ are
bounded. Moreover, $\partial h_1/\partial t$ is bounded under
Assumption~\ref{as:h} and \eqref{eq:sigma}.
Assumption~\ref{as:h} also ensures that
$|\vartheta|\geq\underline{\vartheta}>0$. Hence, $z_2$ is
bounded.
Suppose that $z_1,\ldots,z_{i-1}$ are bounded for
$i\in\{3,\ldots,n\}$. By the recursive definition
\eqref{eq:hi}, the functions
$\partial h_{i-1}/\partial z_j$, $j=1,\ldots,i-2$, and
$\partial h_{i-1}/\partial t$ are finite sums of products
involving mixed partial derivatives of $h_1$ of order at most
$i-1$ and polynomials in $z_2,\ldots,z_{i-1}$. They are
therefore bounded under Assumption~\ref{as:h},
\eqref{eq:sigma}, and the induction hypothesis that $z_1,\ldots,z_{i-1}$ are bounded. Since
$h_i$, $h_{i-1}$ are bounded, the
above expression for $z_i$, together with
$|\vartheta|\geq\underline{\vartheta}>0$, implies that $z_i$
is bounded. By induction, all the states
$z_i$, $i=1,\ldots,n$, are bounded, and hence $|Z(t)|$ is
bounded.
Finally, recalling \eqref{eq:Tran-z}, the nonsingularity of
$T_z$, and the boundedness of $v(t)$ established in
Sec.~\ref{sec:signal}, we conclude that
$Y(t)=T_z^{-1}\bigl(Z(t)-T_vv(t)\bigr)$
is bounded.

According to the theory of Volterra integral equations \cite{Yoshida}, considering the backstepping transformation \eqref{eq:contran1a}, \eqref{eq:contran1b} as well as the boundedness and continuity of the kernels given in Appendix \ref{sec:ker},  there exists a bounded and continuous kernel $\mathcal K(x,y)$ such that
\begin{align}
&\left(
  \begin{array}{c}
    z(x,t) \\
    w(x,t) \\
  \end{array}
\right)
=\left(
                           \begin{array}{c}
                             \alpha(x,t) \\
                             \beta(x,t) \\
                           \end{array}
                         \right)+\int_0^x \mathcal K(x,y)\left(
                           \begin{array}{c}
                             \alpha(y,t) \\
                             \beta(y,t) \\
                           \end{array}
                         \right)dy\notag\\&+\mathcal K_1(x)Y(t)+\mathcal K_2(x)v(t)+\left(
                           \begin{array}{c}
                           \bar\varsigma(x,t) \\
                          {\varsigma}(x,t) \\
                           \end{array}
                         \right)\notag\\&+\int_0^x \mathcal K(x,y)\left(
                           \begin{array}{c}
                            \bar\varsigma(y,t) \\
                         {\varsigma}(y,t) \\
                           \end{array}
                         \right)dy\label{eq:inv}
\end{align}
where $\mathcal K(x,y)$ can be computed from the kernels ${\phi},\varphi,\Psi,\Phi$, and
$\mathcal K_1(x)=\left(\gamma(x),\lambda(x)\right)^T+\int_0^x \mathcal K(x,y)\left(\gamma(y),\lambda(y)\right)^Tdy$, 
$\mathcal K_2(x)=\left(\bar\gamma(x),
                             \bar\lambda(x)\right)^T+\int_0^x \mathcal K(x,y)\left(
                             \bar\gamma(y),
                             \bar\lambda(y)
                         \right)^Tdy$.
Applying the Cauchy-Schwarz inequality to the inverse transformation \eqref{eq:inv},  we have that $z[t],w[t]$ are bounded. Besides, it is obtained from the exponential convergence to zero of $h_1$ that $h(e,t)$ converges to zero according to \eqref{eq:h1} and \eqref{eq:sigma} that shows $\sigma(t)\equiv 0$ after a finite time $t=\bar t_0+t_a$. Recalling Assumption \ref{as:h}, we know $e(t)$ converges to zero. Property 1 is obtained.

2) According to the choice of the design parameters $k_i$, $i=1,\ldots,n-1$ in \eqref{eq:kappai}, recalling Lemma \ref{cl:zi0}, we know $h_i>0$ at $t=\bar t_0=\frac{1}{q_2}$. Because $\beta(0,t)\equiv 0$ for $t\ge \bar t_0$ considering the target system \eqref{eq:targ5}--\eqref{eq:targ8}, and the system matrix $A_{\rm h}$ defined in \eqref{eq:Ah} represents a form of high-relative-degree CBFs $h_i$ as presented in \cite{KrsticBement2006}, \cite{2016Nguyen},  we have $h_i>0$ for $t\ge \bar t_0$. Based on this, we prove the safety in cases a)--c) as follows:\\
a) When the initial safe condition in Definition \ref{def:1} is satisfied, we know from the fact $h(e(\bar t_0),\bar t_0)>0$ that $\sigma(t)\equiv0$ according to \eqref{eq:sigma}. Therefore, it is obtained from $h_1>0$ for $t\ge \bar t_0$ and \eqref{eq:h1} that $h(e(t),t)>0$ for $t\ge \bar t_0$. Thus, together with fact that $h(e(t),t)\ge 0$ holds on $t\in[t_0,\bar t_0]$; ensured by the initial safe condition in Definition \ref{def:1}, we have that $h(e(t),t)\ge 0$ holds on $t\in[t_0,\infty)$.\\
b) When the safe initial conditions in Definition \ref{def:1} is not satisfied, if $h(e(\bar t_0),\bar t_0)>0$, through the same process with the proof in a), we have $h(e(t),t)\ge 0$ holds on $t\in[\bar t_0,\infty)$.\\
c) When the safe initial conditions in Definition \ref{def:1} is not satisfied, if $h(e(\bar t_0),\bar t_0)\le 0$, it is followed that $\sigma(t)\equiv 0$ after a finite time $t=\bar t_0+t_a$ according to \eqref{eq:sigma} where $t_a>0$ is a free design parameter.  We then obtain from the fact $h_1>0$ for $t\ge \bar t_0$ and \eqref{eq:h1} that $h(e(t),t)\ge0$ for $t\ge \bar t_0+t_a$.
Property 2 is thus obtained.
The proof of this theorem is complete.
\end{proof}
In this section, we have proposed the nominal control design based on a fully known model. In practice, the full states, and also the external disturbances, are always inaccessible. Therefore, the next section develops an output-feedback safe controller based only on available measurements.
\section{Unmeasured States and Unknown Disturbances}
The output-feedback design under unmeasured states and unknown disturbances consists of three steps. First, an extended observer is designed to estimate the distributed states and exogenous signals in Sec. \ref{sec:observer}. Next, explicit bounds on the observer errors are derived in Sec. \ref{sec:observererror}. Finally, these bounds are incorporated into the controller design to compensate for the estimation uncertainty and establish the robust output-feedback safe regulator in Sec. \ref{sec:of}.
\subsection{Extended observer design}\label{sec:observer}
We use the measurements $y_1(t)=C_1Y(t)$, $z(1,t)$, together with the known reference signal $r(t)$, to build an extended observer:
\begin{align}
\dot{\hat v}_r&=S_r\hat v_r(t)+ L_r (r(t)-\bar P_r \hat v_r(t)),\label{eq:ob0}\\
\dot {\hat v}_d&=S_d \hat v_d(t)+L_d (z(1,t)-\hat z(1,t)),\label{eq:ob1}\\
\dot {\hat Y}(t) &= A \hat Y(t) + B\hat w(0,t)+\bar G_1\hat v_d(t)+L_y(y_1(t)-C_1\hat Y(t))\notag\\&\quad+L_0(z(1,t)-\hat z(1,t)),\\
{\hat z_t}(x,t) &=  - q_1{\hat z_x}(x,t)+{d_1}\hat w(x,t)+\bar G_2(x)\hat v_d(t)\notag\\&\quad+L_1(x)(z(1,t)-\hat z(1,t)),\label{eq:ob2}\\
\hat {w_t}(x,t) &= q_2\hat {w_x}(x,t)+{d_2}\hat z(x,t)+\bar G_3(x)\hat v_d(t)\notag\\&\quad+L_2(x)(z(1,t)-\hat z(1,t)),\label{eq:ob3}\\
\hat z(0,t) &=p\hat w(0,t)+C \hat Y(t)+\bar G_4\hat v_d(t),\\
\hat w(1,t) &=  qz(1,t)+U(t)+\bar G_5\hat v_d(t),\label{eq:ob6}
\end{align}
where $[\hat v_r,\hat v_d]$ is the estimate of external signals $v^T=[v_r, v_d]$, and $\hat z$, $\hat w$, $\hat Y$ are observer states of the plant, and where $\bar G_i=G_i\bar P_d$, $i=2,3,4,5$. The observer gain $L_y$ is selected such that $A-L_yC_1$ is Hurwitz, considering that  $(A, C_1)$ is observable in Assumption \ref{as:ob}. Other observer gains $L_0$, $L_r$, $L_d$, $L_1(x)$, $L_2(x)$ will be defined later.
Define the estimation error states
\begin{align}
(\tilde v_r, \tilde v_d, \tilde z, \tilde w, \tilde Y)=(v_r, v_d, z, w, Y)-(\hat v_r, \hat v_d, \hat z, \hat w, \hat Y),\label{eq:oberror}
\end{align}
we obtain the observer error system:
\begin{align}
\dot{\tilde v}_r&=(S_r-L_r\bar P_r)\tilde v_r,\label{eq:tildevr}\\
\dot {\tilde v}_d&=S_d \tilde v_d(t)-L_d \tilde z(1,t),\label{eq:tildevd}\\
\dot {\tilde Y}(t) &= (A-L_yC_1) \tilde Y(t) + B\tilde w(0,t)+\bar G_1\tilde v_d(t)-L_0\tilde z(1,t)\notag\\
{\tilde z_t}(x,t) &=  - q_1{\tilde z_x}(x,t)+{d_1}\tilde w(x,t)\notag\\&\quad+\bar G_2\tilde v_ d(t)-L_1(x)\tilde z(1,t),\label{eq:tildea}\\
\tilde {w_t}(x,t) &= q_2\tilde {w_x}(x,t)+{d_2}\tilde z(x,t)\notag\\&\quad-L_2(x)\tilde z(1,t)+\bar G_3\tilde v_ d(t),\label{eq:tildeb}\\
\tilde z(0,t) &=p\tilde w(0,t)+C\tilde Y(t)+\bar G_4\tilde v_ d(t),\label{eq:tildec}\\
\tilde w(1,t) &=  \bar G_5\tilde v_ d(t),\label{eq:tildew1}
\end{align}
recalling \eqref{eq:o1}--\eqref{eq:o5}, \eqref{eq:vrvd}--\eqref{eq:r}. 
\subsubsection{First transformation for the observer error system}
Introduce the transformation:
\begin{align}
\tilde z (x,t) =&\tilde \alpha(x,t) - \int_x^1 K^{11}(x,y)\tilde \alpha(y,t)dy\notag\\
& -  \int_x^1 K^{12}(x,y)\tilde \beta(y,t)dy ,\label{eq:otrana1}\\
\tilde w (x,t) =& \tilde \beta(x,t) - \int_x^1 K^{21}(x,y)\tilde \alpha(y,t)dy \notag\\
&-  \int_x^1 K^{22}(x,y)\tilde \beta(y,t)dy\label{eq:otrana2}\\
{\tilde Y(t)=}&{\tilde X(t)-\int_0^1 K_0(x)\tilde\alpha(x,t)dx-\int_0^1 K_1(x)\tilde\beta(x,t)dx}\label{eq:otrana3}
\end{align}
where $K^{11}(x,y)$, $K^{12}(x,y)$, $K^{21}(x,y)$, $K^{22}(x,y)$ on $\{(x,y)|0\le x\le y\le 1\}$ satisfy
$q_1K^{11}_x(x,y)+q_1K^{11}_y(x,y)=d_1K^{21}(x,y)$,
$q_1K^{12}_x(x,y)-q_2K^{12}_y(x,y)=d_1 K^{22}(x,y)$,
$q_2  K^{21}_x(x,y)-q_1K^{21}_y(x,y)=-d_2K^{11}(x,y)$,
$q_2 K^{22}_x(x,y)+q_2K^{22}_y(x,y)=-d_2K^{12}(x,y)$,
$K^{21}(x,x)=\frac{-d_2}{q_1+q_2}$, $K^{12}(x,x)=\frac{d_1}{q_1+q_2}$,
$K^{11}(0,y)=pK^{21}(0,y)-CK_0(y)$, $K^{22}(0,y)=\frac{1}{p}K^{12}(0,y)-\frac{1}{p}CK_1(y)$, with
{$B K^{21}(0,x)-K_0'(x)q_1+(A-L_yC_1)K_0(x)=0$,
$BK^{22}(0,x)+K_1'(x)q_2+(A-L_yC_1)K_1(x)=0$, $B-K_1(0)q_2=0$, 
$K_0(0)=0$},
whose solution can be obtained as the same process in Appendix \ref{sec:ker} (which can be seen clearly by changing the domain from $\{(x,y)|0\le x\le y\le 1\}$ to $\{(x,y)|0\le y\le x\le 1\}$ and comparing it with (B.4)--(B.6) of \cite{supp}). Applying the transformations \eqref{eq:otrana1}--\eqref{eq:otrana3} and choosing the observer gains as
\begin{align}
L_1(x)=&-p_1(x)+ \int_x^1 K^{11}(x,y)p_1(y)dy\notag\\&+ \int_x^1 K^{12}(x,y)p_2(y)dy-q_1K^{11}(x,1),\label{eq:L1}\\
L_2(x)=&-p_2(x)+\int_x^1 K^{21}(x,y)p_1(y)dy\notag\\&+\int_x^1 K^{22}(x,y)p_2(y)dy-q_1K^{21}(x,1),\label{eq:L}
\end{align}
where $p_1(x),p_2(x)$ are to be determined later,
then \eqref{eq:tildevd}--\eqref{eq:tildew1} is converted into
\begin{align}
\dot {\tilde v}_d&=S_d \tilde v_d(t)-L_d \tilde \alpha(1,t), \label{eq:obe1}\\
\dot{\tilde X}(t)&=(A-L_yC_1)\tilde X(t)+\bar K_0\tilde v_d+(\bar P_1-L_0)\tilde \alpha(1,t)\\
\tilde {\alpha_t}(x,t) &= -q_1\tilde {\alpha_x}(x,t)+\bar K_2(x)\tilde v_ d(t)+p_1(x)\tilde \alpha(1,t),\label{eq:obe2}\\
\tilde {\beta_t}(x,t) &= q_2\tilde {\beta_x}(x,t)+\bar K_1(x)\tilde v_ d(t)+p_2(x)\tilde \alpha(1,t),\\
\tilde {\beta}(1,t) &=\bar G_5\tilde v_ d(t),\\
\tilde \alpha(0,t)& =p\tilde \beta(0,t)+C\tilde X(t)+\bar G_4\tilde v_ d(t),\label{eq:obe7}
\end{align}
where $\bar P_1=\int_0^1 K_0(x)p_1(x)dx+\int_0^1 K_1(x)p_2(x)dx-K_0(1)q_1$,
and
where $\bar K_1(x), \bar K_2(x)$ satisfy
$\bar K_1(x)-  \int_x^1 K^{22}(x,y)\bar K_1(y)dy$ $-\int_x^1 K^{21}(x,y)\bar K_2(y)dy-q_2K^{22}(x,1)\bar G_5-\bar G_3(x)=0$, and $\bar K_2(x)- \int_x^1 K^{11}(x,y)\bar K_2(y)dy-  \int_x^1 K^{12}(x,y)\bar K_1(y)dy-q_2K^{12}(x,1)\bar G_5-\bar G_2(x)=0$, and where
$\bar K_0=\int_0^1 K_0(x)\bar K_2(x)dx$ $+\bar G_1+\int_0^1 K_1(x)\bar K_1(x)dx+K_1(1)q_2\bar G_5$.

\subsubsection{Second transformation for the observer error system}
Applying the second transformation:
\begin{align}
\bar\alpha(x,t)=&\tilde\alpha(x,t)-\Lambda(x)\tilde v_d(t),\label{eq:obtranb1}\\
\bar\beta(x,t)=&\tilde\beta(x,t)-\Lambda_1(x)\tilde v_d(t),\label{eq:obtranb2}\\
\tilde D(t)=&\tilde X(t)-\bar\Lambda\tilde v_d(t)\label{eq:obtranb3}
\end{align}
where $\Lambda_1(x)$, $\Lambda(x)$,  $\bar\Lambda$ satisfy
\begin{align}
&q_2\Lambda_1'(x)-\Lambda_1(x)S_d +\bar K_1(x)=0, \label{eq:La1}\\
&\Lambda_1(1)=\bar G_5,\\
&q_1\Lambda'(x)+\Lambda(x)S_d -\bar K_2(x)=0, \\&\Lambda(0)=p\Lambda_1(0)+ \bar G_4+C\bar\Lambda,\label{eq:La4}\\
&(A-L_yC_1)\bar\Lambda+\bar K_0-\bar\Lambda S_d=0.\label{eq:La5}
\end{align}
The solution of \eqref{eq:La1}--\eqref{eq:La5} is easy to obtain through the following process.
Denote by $\psi_i$ the linearly independent eigenvectors of $S_d$ associated with the eigenvalues $v_i$, $i=1,2,\ldots,n_d$. Post-multiplying \eqref{eq:La1}--\eqref{eq:La5} by $\psi_i$, introducing 
\begin{align}
\iota_i=\Lambda\psi_i,~ \iota_{1,i}=\Lambda_1\psi_i,~ \bar\iota_{i}=\bar\Lambda\psi_i, \label{eq:lambi}
\end{align}
we obtain
$q_1\iota'_i(x)+v_i\iota_i(x)-\bar K_2(x)\psi_i=0$, $\iota_i(0)=p\Lambda_1(0)\psi_i+ \bar G_4\psi_i+C\bar\Lambda\psi_i$,
$q_2\iota'_{1,i}(x)-v_i\iota_{1,i}(x)+\bar K_1(x)\psi_i=0$, $\iota_{1,i}(1)= \bar G_5\psi_i$, and
$$\bar\iota_i=- ((A-L_yC_1)-v_i I)^{-1}\bar K_0 \psi_i.$$
  \begin{remark}
    {\rm Since $A-L_yC_1$ is Hurwitz, all its eigenvalues lie strictly in the open left half-plane.
On the other hand, the eigenvalues of $S_d$ lie on the imaginary axis, as described in Sec. \ref{sec:signal}.
It follows that
$\sigma(A-L_yC_1)\cap\sigma(S_d)=\varnothing.$
Consequently, the matrix $(A-L_yC_1)-v_i I$ is invertible.}
 \end{remark}
 Therefore, according to \eqref{eq:lambi}, the solution of \eqref{eq:La1}--\eqref{eq:La5} are $\Lambda=[\iota_1,\cdots,\iota_{n_d}]\bar V^{-1}$, $\Lambda_1=[\iota_{1,1},\cdots,\iota_{1,n_d}]\bar V^{-1}$, $\bar\Lambda=[\bar\iota_{1},\cdots,\bar\iota_{n_d}]\bar V^{-1}$,  where $\bar V=[\psi_1,\cdots,\psi_{n_d}]$, and where $\iota_i$ and $\iota_{1,i}$ are obtained as
 \begin{align}\label{eq:lambdai}
&\iota_i(x)
=
e^{-\frac{v_i}{q_1}x}\bigg(
pe^{\frac{-v_i}{q_2}}\bar G_5\psi_i
+\frac{p}{q_2}\int_{0}^{1}e^{\frac{-v_i}{q_2}\xi}\bar K_1(\xi)\psi_id\xi
\notag\\&+\bar G_4\psi_i
-C\big((A-L_yC_1)-v_i I\big)^{-1}\bar K_0\psi_i\bigg)\notag\\&
+\frac{1}{q_1}\int_{0}^{x}e^{-\frac{v_i}{q_1}(x-\xi)}\bar K_2(\xi)\psi_id\xi \end{align}
and $\iota_{1,i}(x)
=
e^{\frac{-v_i}{q_2}(1-x)}\bar G_5\psi_i
-\frac{1}{q_2}\int_{x}^{1} e^{\frac{-v_i}{q_2}(\xi-x)}\bar K_1(\xi)\psi_id\xi.
$ 
\subsubsection{Target observer error system}
Defining $p_1(x)$, $p_2(x)$ in the observer gains \eqref{eq:L1}, \eqref{eq:L} are
$$p_1(x)=-\Lambda(x)L_d,~~p_2(x)=-\Lambda_1(x)L_d$$ and choosing $L_0$  as 
\begin{align}
    L_0=\bar\Lambda L_d +\bar P_1,\label{eq:L0}
\end{align} through the second transformation \eqref{eq:obtranb1}--\eqref{eq:obtranb3},
then \eqref{eq:obe1}--\eqref{eq:obe7} is converted into 
\begin{align}
\dot {\tilde v}_d&=(S_d-L_d\Lambda(1)) \tilde v_d(t)-L_d \bar \alpha(1,t),\label{eq:targobe1}\\
\dot{\tilde D}(t)&=(A-L_yC_1)\tilde D(t)\\
\bar {\alpha}_t(x,t) &= -q_1\bar {\alpha}_x(x,t),\label{eq:targobe3}\\
\bar {\beta}_t(x,t) &= q_2\bar {\beta}_x(x,t),\label{eq:targobe4}\\
\bar {\beta}(1,t) &=0,\\
\bar \alpha(0,t)& =p\bar \beta(0,t)+C\tilde D(t).\label{eq:targobe8}
\end{align}
The final target observer error system we arrive at is \eqref{eq:targobe1}--\eqref{eq:targobe8} with \eqref{eq:tildevr}. 
The observer gain $L_r$ is selected such that $S_r-L_r\bar P_r $ Hurwitz, considering $(S_r,\bar P_r)$ is observable in Assumption \ref{as:obex}. Another observer gain $L_d$ is chosen such that $S_d-L_d\Lambda(1)$ is Hurwitz, and the observability of the pair $(S_d,\Lambda(1))$ is shown in the following lemma.
\begin{lema}\label{eq:lemaob}
  The pair $(S_d,\Lambda(1))$ is observable if and only if $\tilde N(v_i)\bar P_d\psi_i\neq 0$, $i=1,\cdots,n_d$, where $\tilde N(s)$ is the numerator of the transfer matrix from $d(t)$ to the measurement $z(1,t)$.
 \end{lema}
 \begin{proof} Applying the same transformation of \eqref{eq:otrana1}--\eqref{eq:otrana3} (replacing $\tilde z, \tilde w, \tilde Y, \tilde \alpha, \tilde \beta, \tilde X$ with $z, w, Y, \check \alpha, \check \beta, \check X$) into the plant \eqref{eq:o1}--\eqref{eq:o5}, we obtain
$\dot{\check X}(t)
=(A-L_yC_1)\check X(t)-q_1K_0(1)\check\alpha(1,t)+\bar K_0v_d(t),$
$\check\alpha_t(x,t)
=-q_1\check\alpha_x(x,t)-q_1K^{11}(x,1)\check\alpha(1,t)+\bar K_2(x)v_d(t),$
$\check\beta_t(x,t)
=q_2\check\beta_x(x,t)-q_1K^{21}(x,1)\check\alpha(1,t)+\bar K_1(x)v_d(t),$
$\check\beta(1,t)
=q\check\alpha(1,t)+\bar G_5v_d(t),$
$\check\alpha(0,t)
=p\check\beta(0,t)+C\check X(t)+\bar G_4v_d(t).$
Rewriting it in the frequency domain, considering $z(1,t)=\check{\alpha}(1,t)$, the transfer function from $v_d(t)$ to $z(1,t)$ is obtained as
$\frac{z(1,s)}{v_d(s)}
=
\frac{N(s)}{D(s)}$,
where
$D(s)=1+\int_0^1e^{-(s/q_1)(1-\xi)}K^{11}(\xi,1)d\xi-pqe^{-s\left(\frac{1}{q_1}+\frac{1}{q_2}\right)}
+e^{-s/q_1}p\int_0^1\frac{q_1}{q_2}e^{-(s/q_2)\xi}K^{21}(\xi,1)d\xi+e^{-s/q_1}q_1C\big(sI-(A-L_yC_1)\big)^{-1}K_0(1),$ and where
\begin{align}\label{eq:N}
&N(s)=
e^{-s/q_1}C\big(sI-(A-L_yC_1)\big)^{-1}\bar K_0\notag\\
&+
e^{-s/q_1}\Big(p\big(\bar G_5 e^{-s/q_2}+\Psi_1(s)\big)+\bar G_4\Big)
+\Phi_2(s),
\end{align}
with
$
\Phi_2(s)=
\frac{1}{q_1}\int_{0}^{1}
e^{-\frac{s}{q_1}(1-\xi)}\bar K_2(\xi)d\xi
$ and
$
\Psi_1(s)
=\frac{1}{q_2}\int_{0}^{1}e^{-\frac{s}{q_2}\xi}\bar K_1(\xi)d\xi.
$ Replacing all $\bar G_i$ involved in $N(s)$ \eqref{eq:N} with $G_i\bar P_d$, we have $N(s)=\tilde N(s)\bar P_d$, where
    $\tilde N(s)=\frac{z(1,s)}{d(s)}$ is the transfer function from $d(t)$ to $z(1,t)$ considering $d(s)=\bar P_d v_d(s)$.  
By the PBH (Popov–Belevitch–Hautus) observability test, the pair $(S_d,\Lambda(1))$ is observable if and only if $\Lambda(1)\psi_i \neq 0$, i.e., $\iota_i(1)\neq 0$ according to \eqref{eq:lambi}. That is,
$\iota_i(1)=N(v_i)\psi_i=\tilde N(v_i)\bar P_d\psi_i\neq 0$ recalling \eqref{eq:lambdai}.
 \end{proof}
 \begin{remark}
 {\rm Lemma \ref{eq:lemaob} shows the observability condition for the disturbance observer, i.e., all eigenmodes of the disturbance model $\dot v_d(t)=S_d v_d(t)$ can be transferred from $d(t)$ to the measurement $z(1,t)$ in steady state, which allows the disturbance observer to reconstruct the state $v_d(t)$ from this measurement. Note that $\bar P_d\psi_i\neq 0$ is ensured by Assumption \ref{as:obex} according to PBH observability test.}\end{remark}
 \subsection{Result of the extended observer}\label{sec:observererror}
Defining the observer error norm
\begin{align}
\tilde\Omega_e=(\|\tilde z(\cdot,t)\|^2+\|\tilde w(\cdot,t)\|^2+|\tilde v|^2+|\tilde Y|^2)^{\frac{1}{2}},\label{eq:normobe}
\end{align}
whose upper bound is estimated in the following lemma, which will be used in building the output-feedback safe controller.
\begin{assm}\label{as:ob1}
    Observability condition in Lemma \ref{eq:lemaob} holds, i.e., $\tilde N(v_i)\bar P_d\psi_i\neq 0$, $i=1,\cdots,n_d$. 
\end{assm}
\begin{lema}\label{eq:lemob}
For initial data $\hat w[t_0]\in L^\infty(0,1)$, $\hat z[t_0]\in L^\infty(0,1)$, $\hat Y(t_0)\in \mathbb R^n$, $\hat v_d(t_0)\in \mathbb R^{n_d}$, $\hat v_r(t_0)\in \mathbb R^{n_r}$,  under Assumption \ref{as:ob1}, with the observer \eqref{eq:ob0}--\eqref{eq:ob6} where the observer gains $L_1(x)$, $L_2(x)$, $L_0$ are defined by \eqref{eq:L1}, \eqref{eq:L}, \eqref{eq:L0}, and $L_d$, $L_r$, $L_y$ satisfy that $S_d-L_d\Lambda(1)$, $S_r-L_r\bar P_r$, and $A-L_yC_1$ are Hurwitz, respectively, the observer errors in the system \eqref{eq:tildevr}--\eqref{eq:tildew1} are exponentially convergent to zero in the sense that
\begin{align}
\tilde\Omega_e(t)\le \Upsilon_2\tilde\Omega_e(t_0) e^{-\sigma_r (t-t_0)},~~~t\in[t_0,\infty)\label{eq:ezw3}
\end{align}
where $\sigma_r>0$ depends on the Hurwitz matrices $S_d-L_d\Lambda(1)$, $S_r-L_r\bar P_r$, $A-L_yC_1$, and where $\Upsilon_2$ depends on the kernels in the transformations \eqref{eq:otrana1}--\eqref{eq:otrana3}, \eqref{eq:obtranb1}--\eqref{eq:obtranb3}.
\end{lema}
\begin{proof}
The estimate \eqref{eq:ezw3} can be obtained by the following Lyapunov analysis. 
Define
\begin{align}
V(t)=&\tilde v^T P_1 \tilde v+ {r_a}(\tilde D^T P_2 \tilde D+ \frac{1}{2}\int_0^1\bar \beta(x,t)^2e^{x}dx)\notag\\&+ \frac{r_c}{2}\int_0^1 \bar \alpha(x,t)^2e^{-x}dx\label{eq:V}
\end{align}
where the positive constants $r_a,r_c$ are to be determined later. The positive definite matrix $P_1= {P_1}^T$ is the solution to the
Lyapunov equation $A_{\rm v}^TP_1+P_1A_{\rm v}=-Q_1$ for some $Q_1={Q_1}^T>0$, where
$A_{\rm v}=[S_r-L_r\bar P_r,0;0,S_d-L_d\Lambda(1)]$.
The positive definite matrix $P_2= {P_2}^T$ is the solution to the
Lyapunov equation $(A-L_yC_1)^TP_2+P_2(A-L_yC_1)=-Q_2$ for some $Q_2={Q_2}^T>0$.
Defining $\tilde\Omega_T(t)=(\|\bar \alpha(\cdot,t)\|^2+\|\bar \beta(\cdot,t)\|^2+|\tilde v|^2+|\tilde D|^2)^{\frac{1}{2}}$, we have
\begin{align}
\xi_1\tilde\Omega_T(t)^2\le V(t)\le \xi_2\tilde\Omega_T(t)^2\label{eq:Xibound}
\end{align}
for some positive $\xi_1$ and $\xi_2$.
Taking the derivative of \eqref{eq:V} along \eqref{eq:targobe1}--\eqref{eq:targobe8} with \eqref{eq:tildevr}, applying Young's and Cauchy-Schwarz inequalities, we get
 \begin{align}
&\dot V(t)\le - \frac{1}{2}\lambda_{\rm min}(Q_1){ {
 \tilde v(t)} ^2} - \bigg(\frac{{q_1} {r_c}}{2} - \frac{{2|P_1B_d|^2}}{\lambda_{\rm min}(Q_1)}\bigg)\bar \alpha(1,t) ^2\notag\\&-\frac{r_c}{2}q_1\int_0^1 {{}{e^{ - x}}{\bar\alpha}{{(x,t)}}^2} dx-\frac{{r_{a}}q_2}{2}\int_0^1 {e^{ x}}{\bar\beta}{{(x,t)}}^2 dx\notag\\&-(r_a-2r_cq_1|C|^2)\lambda_{\rm min}(Q_2)|\tilde D|^2\notag\\&-(q_2r_a-2p^2r_cq_1)\bar\beta(0,t)^2.
\label{eq:dLy}
 \end{align}
Choosing
$r_c>\frac{{4|P_1B_d|^2}}{q_1\lambda_{\rm min}(Q_1)}$,
$r_a>2r_c\max\{2q_1|C|^2,p^2\frac{q_1}{q_2}\}$,
where $B_d=[0,L_d]^T$, we arrive at
$\dot V\le-\sigma_v V(t)$,
where $\sigma_v=\frac{1}{2\xi_2}\min\{\lambda_{\rm min}(Q_1),{r_c}q_1e^{-1},{{r_{a}}q_2},{r_a}\lambda_{\rm min}(Q_2)\}$.
Therefore,
$\tilde\Omega_T(t)\le \sqrt{\frac{\xi_2}{\xi_1}}\tilde\Omega_T(t_0)e^{-\sigma_r (t-t_0)}$,
where $\sigma_r=\frac{1}{2}\sigma_v$. Estimate \eqref{eq:ezw3} is thus obtained by recalling the transformations \eqref{eq:otrana1}--\eqref{eq:otrana3}, \eqref{eq:obtranb1}--\eqref{eq:obtranb3} and applying Cauchy-Schwarz inequality.
\end{proof}
\subsection{Output-feedback safe control}\label{sec:of}
For the observer-based output-feedback control in this section, we require the following assumptions. Assumptions \ref{as:bound}, \ref{as:boundv} indicate that the bounds of the initial values of the plant states and external signals are known, but arbitrary.
\begin{assm}\label{as:bound}
The initial plant states satisfy
$(Y(t_0),z[t_0],$ $w[t_0])\in\mathcal D$,
where
$\mathcal D=
\{
(\ell_1,\ell_2,\ell_3):
\ell_1\in\mathbb R^{n}, \ell_2,\ell_3:[0,1]\rightarrow\mathbb R$$,
\underline Y\le \ell_1\le \overline Y,
\underline z_0(x)\le \ell_2(x)\le \overline z_0(x),
\underline w_0(x)\le \ell_3(x)\le \overline w_0(x)
\}.$
The functions $\underline z_0(x)$, $\overline z_0(x)$, $\underline w_0(x)$, $\overline w_0(x)$, together with the constant vectors $\underline Y$ and $\overline Y$, are known but arbitrary.
\end{assm}
\begin{assm}\label{as:boundv}
The initial values of exogenous signals satisfy $v(t_0)\in \mathcal D_v$, where $\mathcal D_v=\{\ell\in\mathbb R^{n_{v}}|\underline v\le \ell\le \overline v\}$, and the constant vectors $\underline v,\overline v\in\mathbb R^{n_{v}}$ are known and arbitrary.
\end{assm}
\begin{assm}\label{as:asign}
The sign of $h(e(\bar t_0),\bar t_0)$ is known.
\end{assm}For Assumption \ref{as:asign}, the exact value of $h(e(\bar t_0),\bar t_0)$ is not required but its sign that will be used in determining the form of $\sigma$ in \eqref{eq:sigma}, which can be judged based on the known bounds of the initial values in Assumptions \ref{as:bound}, \ref{as:boundv}.
\begin{assm}\label{as:h1}
The barrier function $h$ ensures that there exists a positive $\xi_e$ such that
\begin{align}
&|g(T_zY(t)+T_vv(t),z[t],w[t],v(t),t)\notag\\&-g(T_z\hat Y(t)+T_v\hat v(t),\hat z[t],\hat w[t],\hat v(t),t)|\notag\\&\le \xi_e (|\tilde Y(t)|+|\tilde v(t)|+\|\tilde z\|+\|\tilde w\|)\label{eq:tildef}
\end{align}
where $g $ is determined by $h$ via \eqref{eq:p1t}, \eqref{eq:f}, \eqref{eq:h1}, \eqref{eq:hi}, and \eqref{eq:hy1}.
\end{assm}
{Assumption \ref{as:h1} imposes an additional condition on the barrier function $h$ compared to Assumption \ref{as:h}, considering the uncertainties of the plant state and the external disturbances in the output-feedback case. Assumption \ref{as:h1} can also be verified easily when planning the barrier function $h$ using \eqref{eq:f}, \eqref{eq:h1}, \eqref{eq:hi}.}

Considering the uncertainties, the design parameters $k_i$ are now chosen as
\begin{align}
k_i&>\max_{{(\ell_1,\ell_2,\ell_3)\in \mathcal D, \ell_4\in \mathcal D_v}}\{0,\acute k_i(\ell_1,\ell_2,\ell_3,\ell_4)\}, \label{eq:ko}
\end{align}
$\forall i=1,2,\cdots,n-1$, where $\acute k_i$ are defined in
\eqref{eq:kappai}, and where $\mathcal D$, $\mathcal D_v$ given in Assumptions \ref{as:bound}, \ref{as:boundv}. Besides, there also exists conservatism in the choice of $\sigma(t)$, which arises from using known bounds to estimate the lower bound of $h(e(\bar t_0),\bar t_0)$, denoted as
$\underline h(e(\bar t_0),\bar t_0)=\inf_{(\ell_1,\ell_2,\ell_3)\in \mathcal D, \ell_4\in \mathcal D_v}\mathcal P(\ell_1,\ell_2(\cdot),\ell_3(\cdot),\ell_4,\frac{1}{q_2})$,
where the function $\mathcal P$ is defined in \eqref{eq:mathp}. The lower bound $\underline h(e(\bar t_0),\bar t_0)$ is used in place of $h(e(\bar t_0),\bar t_0)$ in the coefficient of the function  $\sigma(t)$ \eqref{eq:sigma} (the sign of $h(e(\bar t_0),\bar t_0)$ used in judging the case in \eqref{eq:sigma} is known according to Assumption \ref{as:asign}).
Design the output-feedback safe controller as
\begin{align}
U_f=\hat U+{\rm sign}(\vartheta(e(t_0),t_0))M_ce^{-\sigma_r (t-t_0)}\label{eq:U2}
\end{align}
where 
\begin{align}
   &\hat U=-qz(1,t)+\int_0^1 {\Psi}(1,y)\hat z(y,t)dy+\int_0^1 {\Phi}(1,y)\hat w(y,t)dy\notag\\&+\lambda(1)\hat Y(t)- (\acute G_5-\bar\lambda(1)) {\hat v}(t)+\hat\varsigma(1,t)\label{eq:hU}
\end{align}
 is obtained by replacing the unmeasurable states $z(x,t)$,  $w(x,t)$, $Y(t)$ and the external disturbances $v(t)$ in the nominal control law \eqref{eq:U1} with their estimates, and $\hat\varsigma(1,t)=g(T_z\hat Y(t)+T_v\hat v(t),\hat z[t],\hat w[t],\hat v(t),t)$, and where the sign is defined by ${\rm sign}(u)= \begin{cases}
1 &  u\ge 0
  \\
-1 & u<0
\end{cases}$
          and the constant $M_c$ is given by
          \begin{align}M_c=& 4\max\big\{\|\Psi(1,y)\|,\|\Phi(1,y)\|,|\lambda(1)|,\notag\\&|\acute G_5-\bar\lambda(1)|,\xi_e\big\}\Upsilon_2\overline\Omega.
\label{eq:rhoe}
          \end{align}
          The constant $\xi_e$ is defined in Assumption \ref{as:h1}, $\Upsilon_2$ is the overshoot coefficient in Lemma \ref{eq:lemob}, $\Psi$, $\Phi$ and $\lambda$ are bounded kernels in \eqref{eq:contran1b}, and the constant $\overline\Omega$, which is the upper bound of the initial observer error norm $\tilde\Omega_e(t_0)$, i.e., $\tilde\Omega_e(t_0)\le \overline\Omega$, can be obtained by \eqref{eq:normobe} and the known bounds in Assumptions \ref{as:bound}, \ref{as:boundv}, under the fact that the initial observer states are chosen within the corresponding known bounds. As will be seen clearly later, the function $M_ce^{-\sigma_r t}$ is indeed an estimated upper bound on the absolute value of the error between the state-feedback and output-feedback controllers, which results from the unmeasured states and the external signals. The role of ${\rm sign}(\vartheta(e(t_0),t_0))M_ce^{-\sigma_r t}$ in \eqref{eq:U2} is to tolerate this error by shrinking the original safe set to a subset that converges to the original safe set as time approaches infinity.
\begin{thme}\label{th:th2}
{Assume that the closed-loop system consisting of the plant \eqref{eq:o1}--\eqref{eq:o5}, the observer \eqref{eq:ob0}--\eqref{eq:ob6}, and the controller \eqref{eq:U2} admits a classical solution.} Under Assumptions \ref{as:obex}--\ref{as:h1}, for initial data $w[t_0]\in L^\infty(0,1)$, $z[t_0]\in L^\infty(0,1)$, $Y(t_0)\in \mathbb R^n$, $v(t_0)\in \mathbb R^{n_v}$, and initial observer states $\hat w[t_0]\in L^\infty(0,1)$, $\hat z[t_0]\in L^\infty(0,1)$, $\hat v(t_0)\in \mathbb R^{n_v}$, $\hat Y(t_0)\in \mathbb R^n$ are selected within the known bounds given by Assumptions \ref{as:bound}, \ref{as:boundv}, and for design parameters $k_i$, $i=1,\ldots,n-1$ satisfying \eqref{eq:ko}, and $k_n>0$, the following properties hold:\\
1) Tracking error $e(t)=y_1(t)-r(t)$ is convergent to zero and all states are bounded;\\
2) Safety is ensured in the sense that:\\
a) When the safe initial condition in Definition \ref{def:1} is satisfied, the safety is ensured from $t=t_0$, i.e., $h(e(t),t)\ge 0,~\forall t\ge t_0$;\\
b) When the safe initial condition in Definition \ref{def:1} is not satisfied, and $h(e(\bar t_0),\bar t_0)>0$, the safety is ensured from $t=\bar t_0$, i.e., $h(e(t),t)\ge 0,~\forall t\ge \bar t_0$;\\
c) When the safe initial condition in Definition \ref{def:1} is not satisfied, and $h(e(\bar t_0),\bar t_0)\le0$, the state will return and stay in the safe region no later than a finite time $\bar t_0+t_a$ where the constant $t_a>0$ can be arbitrarily assigned by users, i.e., $h(e(t),t)\ge 0,~\forall t\ge \bar t_0+t_a$.
\end{thme}
\begin{proof}
1) Define difference between the state-feedback control $U$ and the output-feedback control $\hat U$ as $\eta_e(t)$. Recalling  \eqref{eq:hU} and \eqref{eq:U1}, we have
\begin{align}
\eta_e(t)=&U-\hat U\notag\\=&\int_0^1 {\Psi}(1,y)\tilde z(y,t)dy +\int_0^1 {\Phi}(1,y)\tilde w(y,t)dy\notag\\&+\lambda(1)\tilde Y(t)- (\acute G_5-\bar\lambda(1)) \tilde {v}(t)+\tilde \varsigma(1,t),\label{eq:etae}
\end{align} where $\tilde \varsigma(1,t)=\varsigma(1,t)-\hat \varsigma(1,t)$.
Applying the output-feedback controller \eqref{eq:U2},  we have
\begin{align}
\beta(1,t)={\rm sign}(\vartheta(e(t_0),t_0))M_ce^{-\sigma_r (t-t_0)}-\eta_e(t).\label{eq:dothmob}
\end{align}
The target system becomes \eqref{eq:targ5}--\eqref{eq:targ4} with \eqref{eq:dothmob}. 
Recalling \eqref{eq:rhoe}, \eqref{eq:etae}, \eqref{eq:p1t}, and \eqref{eq:tildef} in Assumption \ref{as:h1}, applying Cauchy-Schwarz inequality and Lemma \ref{eq:lemob}, we obtain
\begin{align}
|\eta_e(t)|\le&4\max\big\{\|\Psi(1,y)\|,\|\Phi(1,y)\|,|\lambda(1)|,\notag\\&|\acute G_5-\bar\lambda(1)|,\xi_e\big\}\tilde\Omega_e(t)
\le M_ce^{-\sigma_r (t-t_0)}\label{eq:ee}
\end{align}
where $\tilde\Omega_e(t)$ is defined in \eqref{eq:normobe}.
Therefore, $\beta(1,t)$ in \eqref{eq:dothmob} is exponentially convergent to zero. Through the same process in the proof of Property 1 in Theorem \ref{th:th1}, Property 1 of this theorem is obtained.

2) For Property 2, since $\vartheta(e(t),t)$ is continuous and nonzero for all
$t\geq t_0$, its sign does not change, i.e.,
$
{\rm sgn}(\vartheta(e(t),t))
=
{\rm sgn}\big(\vartheta(e(t_0),t_0)\big),
t\geq t_0.
$
It follows from  \eqref{eq:dothmob}, \eqref{eq:ee} that
$
{\rm sgn}(\vartheta(e(t_0),t_0))\beta(1,t)
\geq 0,
\quad t\geq t_0.
$
Moreover, since $\beta_t=q_2\beta_x$, one has
$
\beta(0,t)=\beta(1,t-\frac{1}{q_2}),
\quad t\geq \bar t_0:=t_0+\frac{1}{q_2},
$
and hence
$
\vartheta(e(t),t)\beta(0,t)\geq 0,
\quad t\geq \bar t_0.
$
Therefore, from the last equation of \eqref{eq:targ5},
$
\dot h_n(t)
=
-k_n h_n(t)
+b\vartheta(e(t),t)\beta(0,t)
\geq -k_n h_n(t)
$,
where $b>0$ and $k_n>0$. The choice of the design parameters $k_i$ in \eqref{eq:ko} is a subset of the ones \eqref{eq:kappai}, which thus ensures $h_i>0$, $i=1,\cdots,n$ at $t=\bar t_0$ according to Lemma \ref{cl:zi0}. Therefore, we have
$
h_n(t)
\geq
h_n(\bar t_0)e^{-k_n(t-\bar t_0)}
>0,
\quad t\geq \bar t_0.
$
For $i=1,\ldots,n-1$, \eqref{eq:targ5} gives
$
\dot h_i(t)=-k_i h_i(t)+h_{i+1}(t).
$
Consequently, starting from $h_n(t)>0$, $t\geq \bar t_0$, recursively applying the comparison
principle, and recalling $h_i(\bar t_0)>0$, yields
$
h_i(t)>0,
\quad t\geq \bar t_0
$, $i=1,\cdots,n$.
Recalling that
$h_1(t)=h(e(t),t)+\sigma(t)
$, the above positivity result gives the desired safety properties. In particular, 
for the safe initial condition in Definition \ref{def:1}, $\sigma(t)\equiv0$, and hence
$h(e(t),t)\geq0$, $t\ge \bar t_0$. Together with Definition \ref{def:1}, the safety is ensured for all $t\ge t_0$. For an initially unsafe condition satisfying
$h(e(\bar t_0),\bar t_0)>0$, the same conclusion holds for all
$t\geq\bar t_0$. Finally, when $h(e(\bar t_0),\bar t_0)\leq0$, the
term $\sigma(t)$ vanishes after the prescribed time $\bar t_0+t_a$, and therefore
$h(e(t),t)\geq0$ thereafter. This completes the proof of Property 2.
\end{proof}
\section{Application in UAV Safe Delivery}
\begin{figure}
\centering
\includegraphics[width=6.2cm]{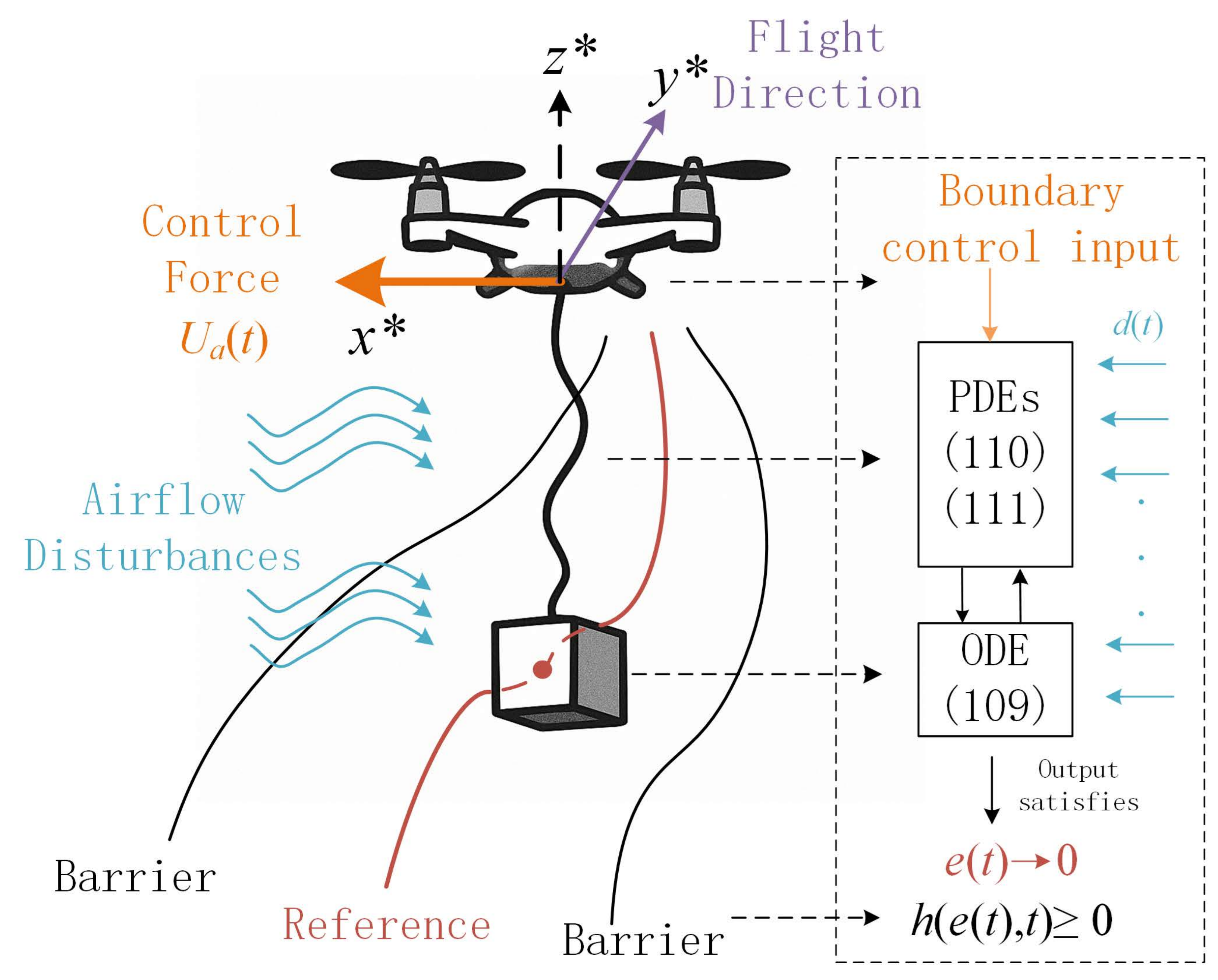}
\caption{UAV delivery with cable-suspended payloads.}
\label{fig:uav}
\end{figure}

In the simulation, we apply our control design to a UAV system with cable-suspended payloads for delivery as shown in Fig. \ref{fig:uav}, aiming to regulate the payload at the lower end of the cable to track a desired reference while avoiding collisions of the payload with surrounding barriers. For simplicity, we consider only one-dimensional motion.
\subsection{Model}\label{sec:simmod}
We adopt the following simplifications in modeling:
a) We neglect the UAV dynamics, assuming that the control force is applied directly at the top of the cable.
b) Assuming the UAV is flying at a predetermined speed along the $y^*$ direction, we only focus on the safe regulation problem in the $x^*$ direction.
Then, according to \cite{J2020delay}, the oscillatory dynamics of the cables are modeled by wave PDEs, while the lumped payload at the cable tip is described by a linear ODE. Additionally, wind disturbances $f_i$, which act throughout the cable-payload system, are also taken into account. Thus, the mathematical model is formulated as follows:
\begin{align}
&M_0\ddot y_1(t) = -d_0\dot y_1 (t)+ T_0u_x(0,t)+f_1(t),\label{eq:5.1}\\
&\rho u_{tt}(x,t) = T_0u_{xx}(x,t)- d_c{u_t}(x,t)+f_2(x,t), \label{eq:5.3}\\
&u(0,t) = y_1(t),~T_0u_x(L,t) = U_a(t)+f_3(t),\label{eq:5.4}
\end{align}
for $x\in[0,L]$. Therein, the function $u(x,t)$ represents the distributed transverse displacement along the cable, while $y_1(t)$ is the transverse displacement of the payload. The constant $T_0=M_0g$ denotes the static tension. The terms $f_i$ denote airflow disturbances, which will be given in detail later. The values of the physical parameters are listed in Tab. \ref{table1}. To evaluate the controller under more demanding conditions, we intentionally introduce anti-damping, making the simulation model open-loop unstable. In this setting, instability sources are present in both PDE and ODE subsystems.
\begin{table}
\centering
\caption{Physical parameters}
\begin{tabular}{lccc}
\hline
Parameters (units)&values\\ \hline
Cable length ${L}$ (m) &1\\
Cable linear density ${\rho}$ (kg/m) &0.5\\
Payload mass ${M_0}$ (kg) &15\\
Gravitational acceleration $g$ (m/s$^{2}$) &9.8\\
Cable material damping coefficient $d_c$ (N$\cdot$s/m) & -1\\
Damping coefficient at payload $d_0$ (N$\cdot$s/m) & -1\\
\hline
\tiny {Note: anti-damping is introduced to increase challenges.}
\end{tabular}
\label{table1}
\end{table}
Applying Riemann transformations
${z}(x,t) = {u_t}(x,t) - \sqrt {\frac{T_0}{\rho}} {u_x}(x,t)$,
${w}(x,t) = {u_t}(x,t) + \sqrt {\frac{T_0}{\rho}} {u_x}(x,t)$
and defining a new variable $Y(t)=[y_1(t),\dot y_1(t)]^T$, \eqref{eq:5.1}--\eqref{eq:5.4} can be rewritten as
\begin{align}
&\dot Y(t) =AY(t) + B{w}(0,t)+G_1d(t),\label{eq:s1}\\
&z(0,t)= pw(0,t)+CY(t),\label{eq:s4}\\
&{z_{t}}(x,t) =  - q_1 {z_{x}}(x,t)+ c{z}(x,t)\notag\\& + c{w}(x,t)+G_2(x)d(t),\label{eq:s2}\\
&{w_{t}}(x,t) = q_2{w_{x}}(x,t) + c{z}(x,t)\notag\\& + c{w}(x,t)+G_3(x)d(t),\label{eq:s3}\\
&w(L,t) = qz(L,t)+G_5d(t)+U(t),\label{eq:s5}
\end{align}
where
$q_1=q_2=\sqrt {\frac{T_0}{\rho}}, c=\frac{{{-d_c}}}{2\rho}, p=-1, q=1$,
$A=\left[0, 1 ;0 , \frac{-d_0}{M_0}-\frac{\sqrt{T_0\rho}}{M_0} \right], B=[0,\frac{\sqrt {T_0\rho}}{M_0}]^T, C=[0,2], U=2\frac{1}{\sqrt{T_0\rho}}U_a(t)$.
After the Riemann transformation in Sec. \ref{sec:simmod}, the airflow disturbances $f_i$ are given by the external disturbances $d(t)=\bar P_d v_d(t)$ \eqref{eq:vd} with $\bar P_d=I_{4\times 4}$ and
$
v_d=[\sin(0.25t),\cos(0.25t),\sin(0.5t),\cos(0.5t)]
$
in the $2\times 2$ hyperbolic PDEs \eqref{eq:s1}--\eqref{eq:s5}, where $G_1=[g_1, g_2]^T$ with $g_1=[0,0,0,0]$, $g_2=[1,1,1,1]$, $G_5=[1,0,1,0]$, $G_2(x)=[x,0,0,0]$, $G_3(x)=[x,0,0,0]$. 
The desired reference is set as
$r(t)=\sin(0.25\pi t)+\cos(0.25\pi t)$, i.e., $\bar P_r=[1,1]$, and
$v_r=[\sin(0.25\pi t),\cos(0.25\pi t)]$
in \eqref{eq:r}.
Therefore, the system matrix of exogenous signal model is $S_d=[0,0.25,0,0;-0.25,0,0,0;0,0,0,0.5;0,0,$ $-0.5,0]$ and $S_r=[0,0.25\pi;-0.25\pi,0]$.
\begin{remark}
{\rm Compared to the plant described by \eqref{eq:o1}--\eqref{eq:o5}, the simulation model represented by \eqref{eq:s1}--\eqref{eq:s5} includes additional terms, specifically $c z(x,t)$ and $c w(x,t)$ in \eqref{eq:s2} and \eqref{eq:s3}, respectively. However, these additions do not alter the state-feedback control design presented in this paper. To be exact, applying state-feedback control, the target system defined by \eqref{eq:targ5} and \eqref{eq:targ8}, which corresponds to the closed-loop system, remains unchanged except for the updates that \eqref{eq:targ4} are modified to $\beta_t = q_2\beta_x + c\beta$ and $\alpha_t = -q_1\alpha_x + c\alpha$. Along their characteristic curves, the additional terms $c\beta$, $c\alpha$ introduce only positive multiplicative factors, which preserve the signs of the transported states and do not alter the characteristic travel times. Therefore, they do not impact the safety and stability results obtained. The corresponding change in the observer-based output-feedback controller is adding the terms $c\hat{z}(x,t)$ and $c\hat{w}(x,t)$ in \eqref{eq:ob2} and \eqref{eq:ob3} of the observer.}
\end{remark}
The initial values are defined as
$w(x,0)=\cos(2\pi x)$, $z(x,0)=\sin(3\pi x)$, $y_2(0)=0$, $\hat w(x,0)=\cos(2\pi x)+0.2$, $\hat z(x,0)=\sin(3\pi x)+0.2$, $\hat Y(0)=[\hat y_1(0),\hat y_2(0)]=[y_1(0), y_2(0)+0.2]^T$, $\hat v_d(0)=v_d(0)+0.2=[0.2,1.2,$ $0.2,1.2]^T$, $\hat v_r(0)=v_r(0)+[0.2,-0.2]=[0.2,0.8]^T$. For safe initialization, we set $y_1(t_0)=8$. For the unsafe initialization case, we set $y_1(t_0)=-1$. The known bounds of the initial values of the unmeasurable states in Assumptions \ref{as:bound}, \ref{as:boundv} are set to $\pm 0.2$ around their true values.
The simulation is conducted using the finite difference method with the space step as $0.05$ and the time step as $0.001$.
The barrier function is set as
$h(e(t),t)=e(t)-3e^{-0.4t}=y_1(t)-r(t)-3e^{-0.4t}$,
where $r(t)$ is the desired reference trajectory of the regulated payload displacement. Consequently, the time-varying safe region is $y_1(t)\ge r(t)+3e^{-0.4t}$. 
\subsection{Controller and simulation results}
The state-feedback safe regulator is obtained from \eqref{eq:U1}, with the control parameters selected as $k_1=0.65$, $k_2=1.4$.
The output-feedback safe regulator $U_f$ is obtained from \eqref{eq:U2}. For both the safe- and unsafe-initialization cases, the control parameters are selected as $k_1=5$, $k_2=8$, $M_c=215$, $\sigma_r=0.35$.
The same observer gains are employed in the safe- and unsafe-initialization cases. The gain $L_y$ is chosen such that the eigenvalues of $A-L_yC_1$ are assigned to $[-0.75,-0.95]$. The gain $L_d$ is selected such that the eigenvalues of $S_d-L_d\Lambda(1)$ are assigned to $[-1.7\pm 0.25 i,-1.55\pm 0.5 i]$. Similarly, the gain $L_r$ is obtained by assigning the eigenvalues of $S_r-L_r\bar P_r$ to $[-0.9\pm \frac{\pi}{4} i]$.
For the unsafe-initialization case, the initial payload displacement is selected as $y_1(0)=-1$. The same output-feedback controller and observer gains are retained, while the recovery function $\sigma(t)$ in \eqref{eq:sigma} is activated with $\epsilon=1.05$, $t_a=1.5$, to drive the output back into the time-varying safe region.

For comparison, an output regulation controller without the safety mechanism is also implemented. It is designed directly using the third backstepping transformation, where the safety-related last terms are removed. The resulting control law is $U_{\rm O}(t)=-qz(1,t)+\int_{0}^{1}\Psi_{\rm O}(1,y)\hat z(y,t)dy+\int_{0}^{1}\Phi_{\rm O}(1,y)\hat w(y,t)dy+\lambda_{\rm O}(1)\hat Y(t)-(\acute G_{5}-\bar{\lambda}_{\rm O}(1))\hat v(t)$, where $\lambda_{\rm O}$ is selected to assign the poles of the target ODE system matrix $A+B\lambda_{\rm O}(0)$ to $[-5,-6]$, and $\Psi_{\rm O}$, $\Phi_{\rm O}$ are obtained in the same manner as $\Psi$ and $\Phi$, except that $\lambda$ is replaced by $\lambda_{\rm O}$. The function $\bar{\lambda}_{\mathrm O}$ is obtained from the same regulator equations as $\bar{\lambda}$, with $\lambda$, $\Psi$, and $\Phi$ replaced by $\lambda_{\mathrm O}$, $\Psi_{\mathrm O}$, and $\Phi_{\mathrm O}$, respectively.
\begin{figure}
\includegraphics[width=8.1cm]{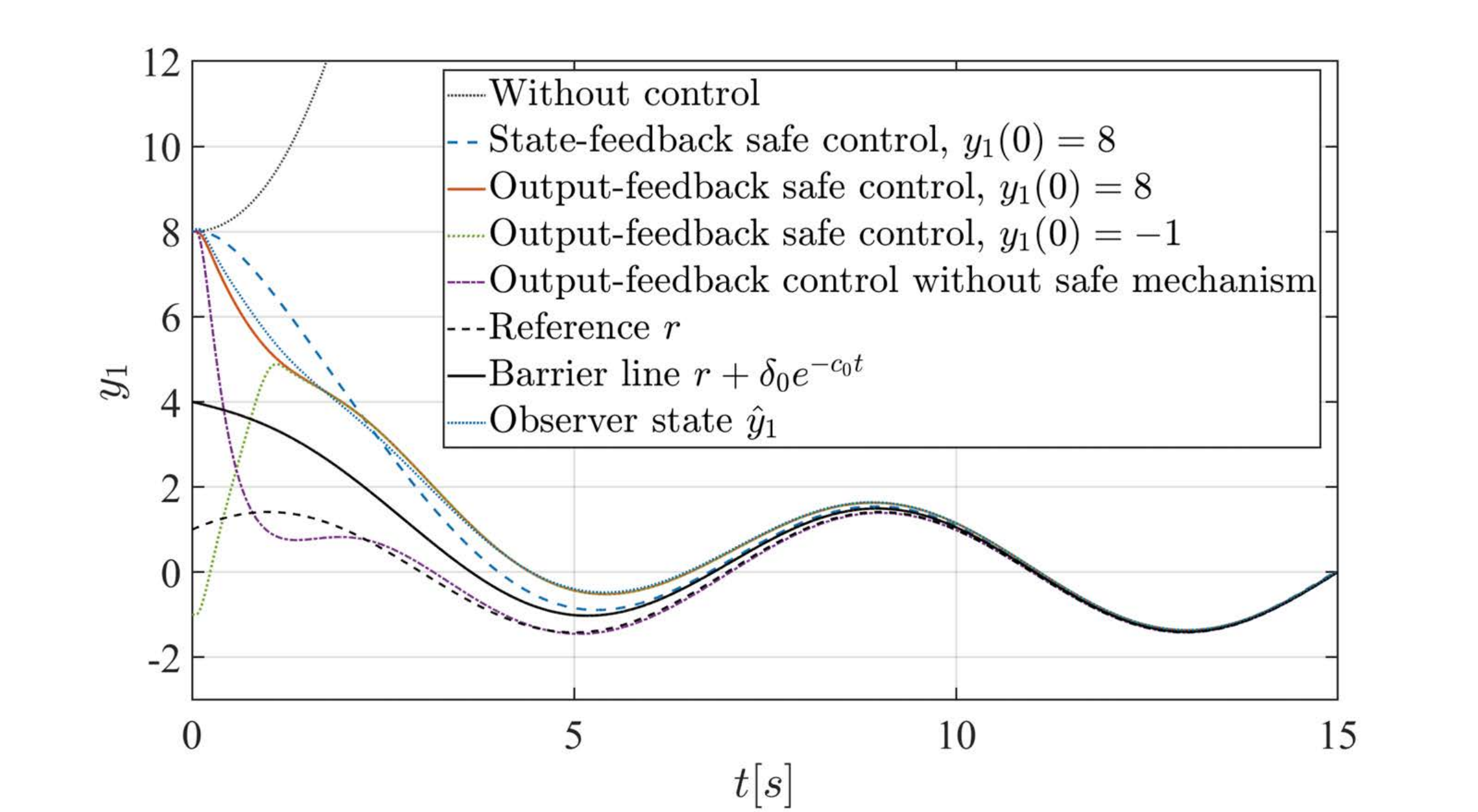}
\caption{Payload displacement $y_1$}
\label{fig:y1}
\end{figure}
\begin{figure}
\includegraphics[width=8.1cm]{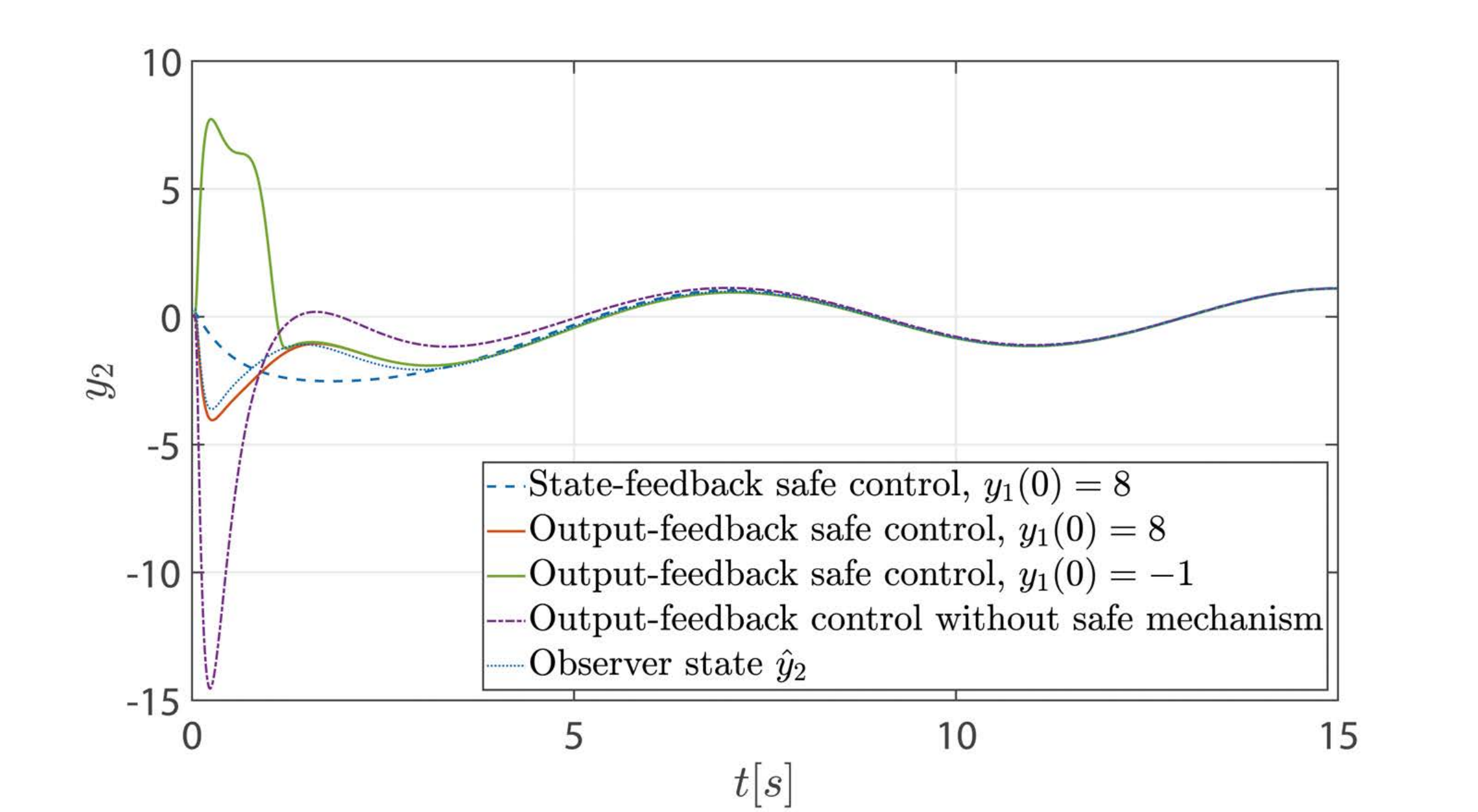}
\caption{Payload velocity $y_2$}
\label{fig:y2}
\end{figure}
\begin{figure}
\begin{minipage}{0.49\linewidth}
  \centerline{\includegraphics[width=4.3cm]{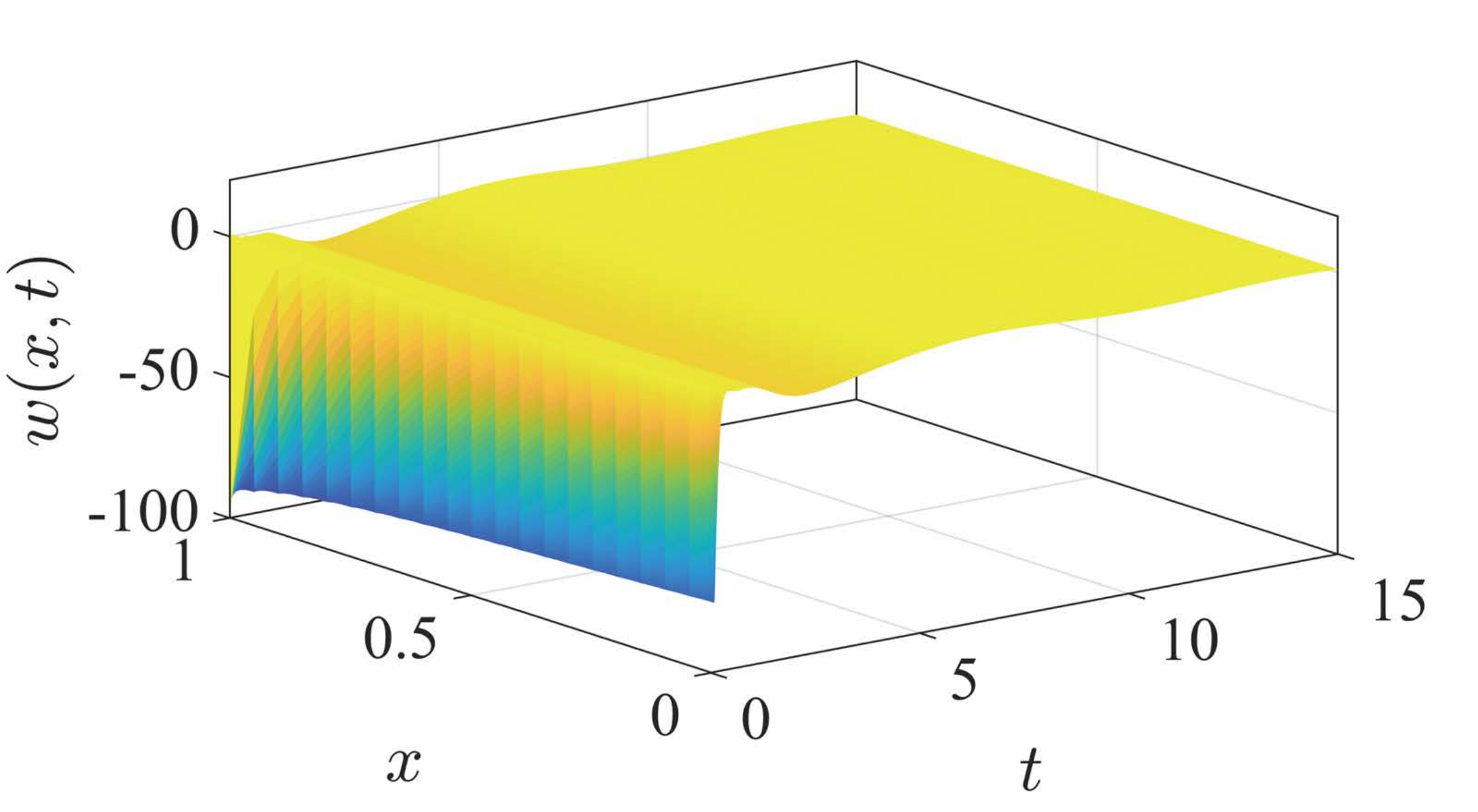}}
  \centerline{(a)  $w(x,t)$ }
\end{minipage}
\hfill
\begin{minipage}{.49\linewidth}
  \centerline{\includegraphics[width=4.3cm]{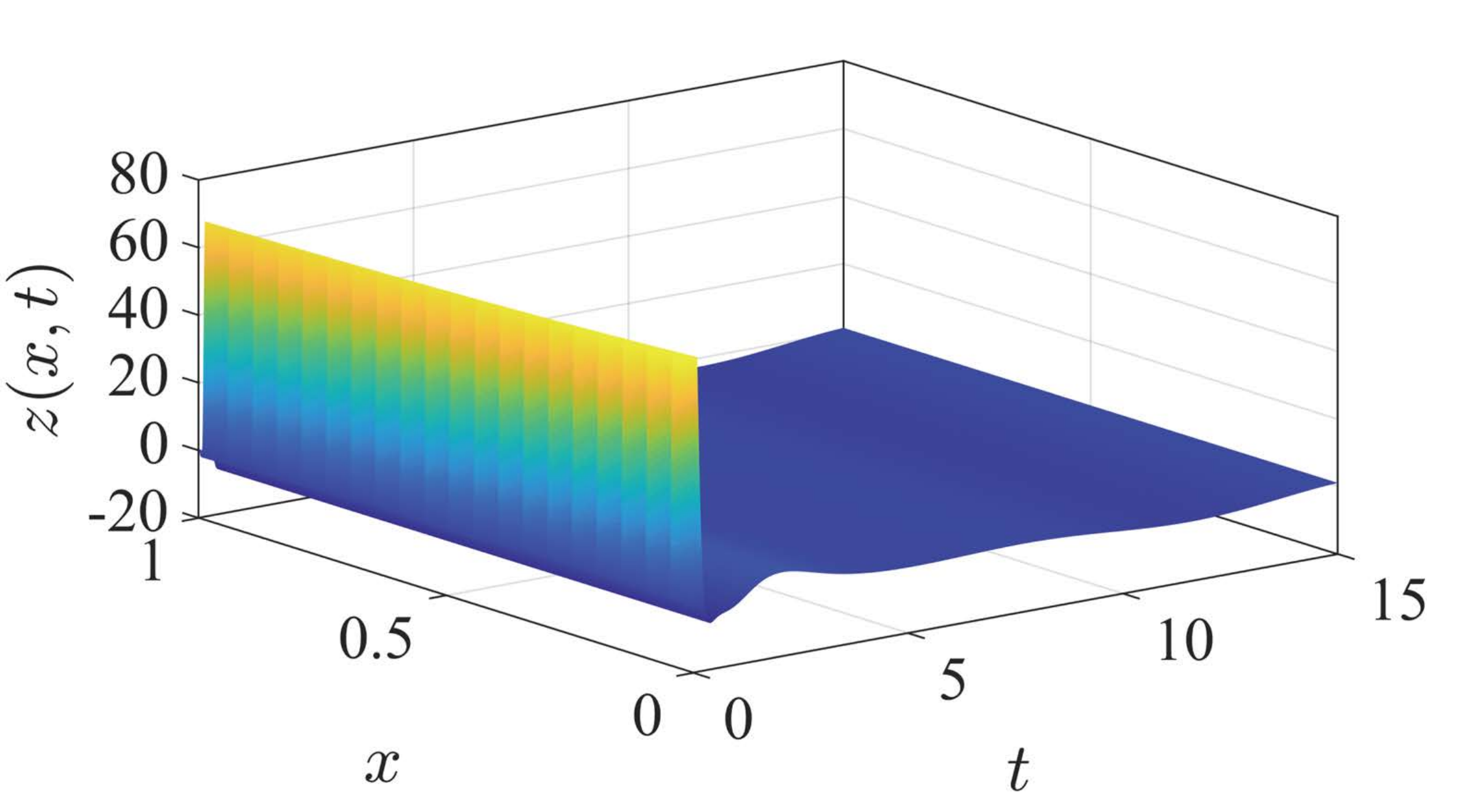}}
  \centerline{(b)  $z(x,t)$}
\end{minipage}
\caption{PDE states with the output-feedback safe control.}
\label{fig:pdestates}
\end{figure}
\begin{figure}
\begin{minipage}{0.49\linewidth}
  \centerline{\includegraphics[width=4.3cm]{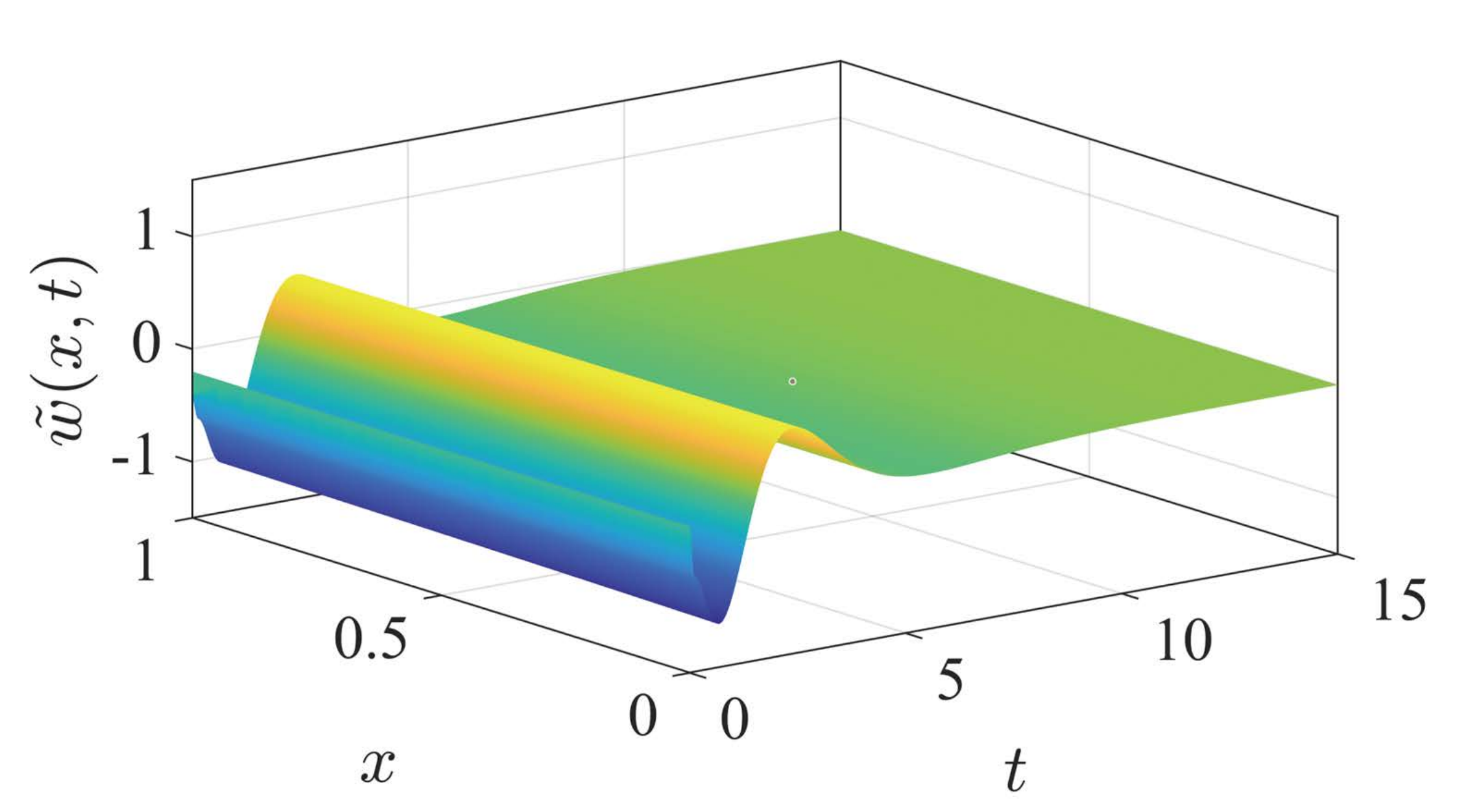}}
  \centerline{(a)  $\tilde w(x,t)$ }
\end{minipage}
\hfill
\begin{minipage}{.49\linewidth}
  \centerline{\includegraphics[width=4.3cm]{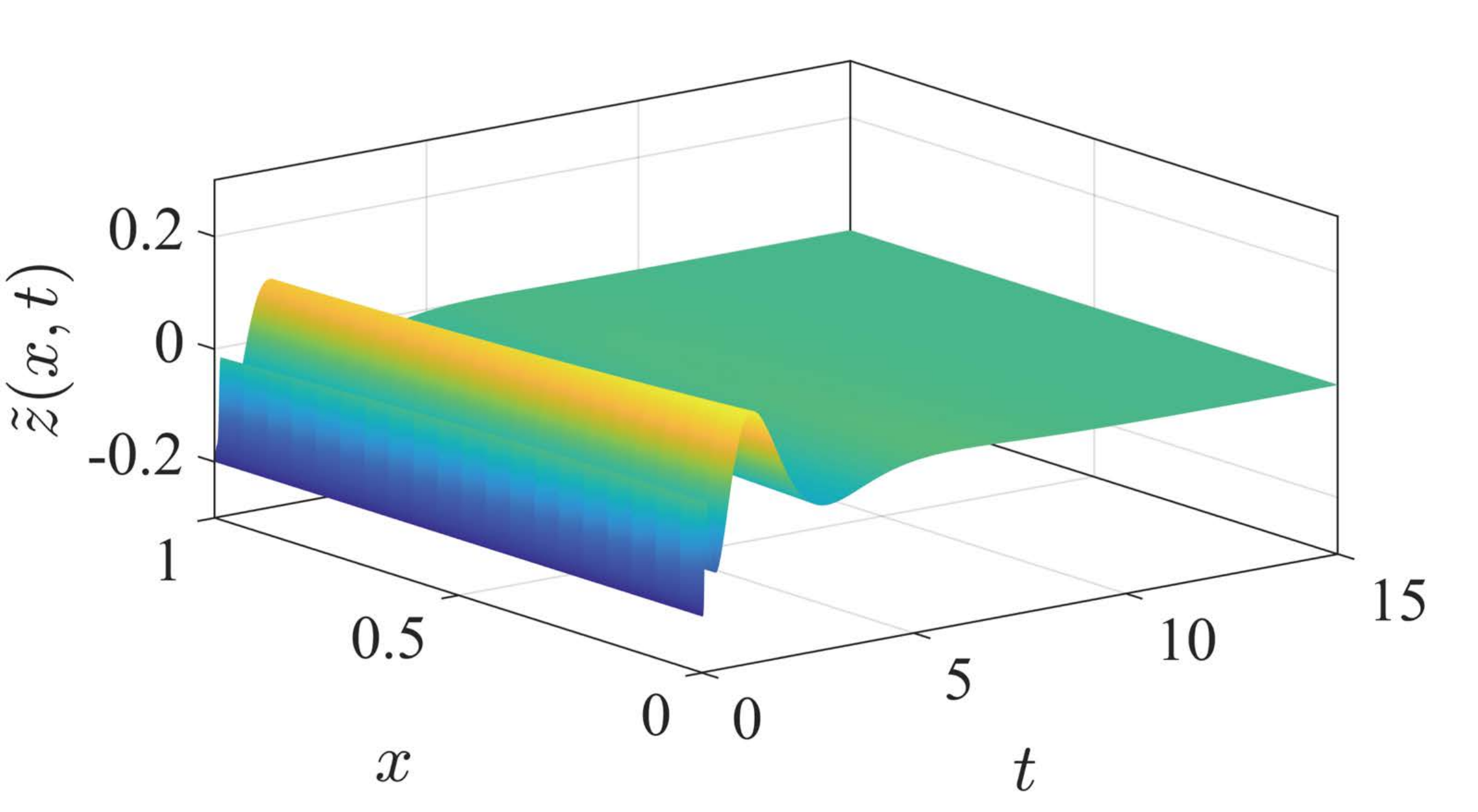}}
  \centerline{(b) $\tilde z(x,t)$}
\end{minipage}
\caption{Observer errors}
\label{fig:ob}
\end{figure}
\begin{figure}
\begin{minipage}{0.49\linewidth}
  \centerline{\includegraphics[width=4.2cm]{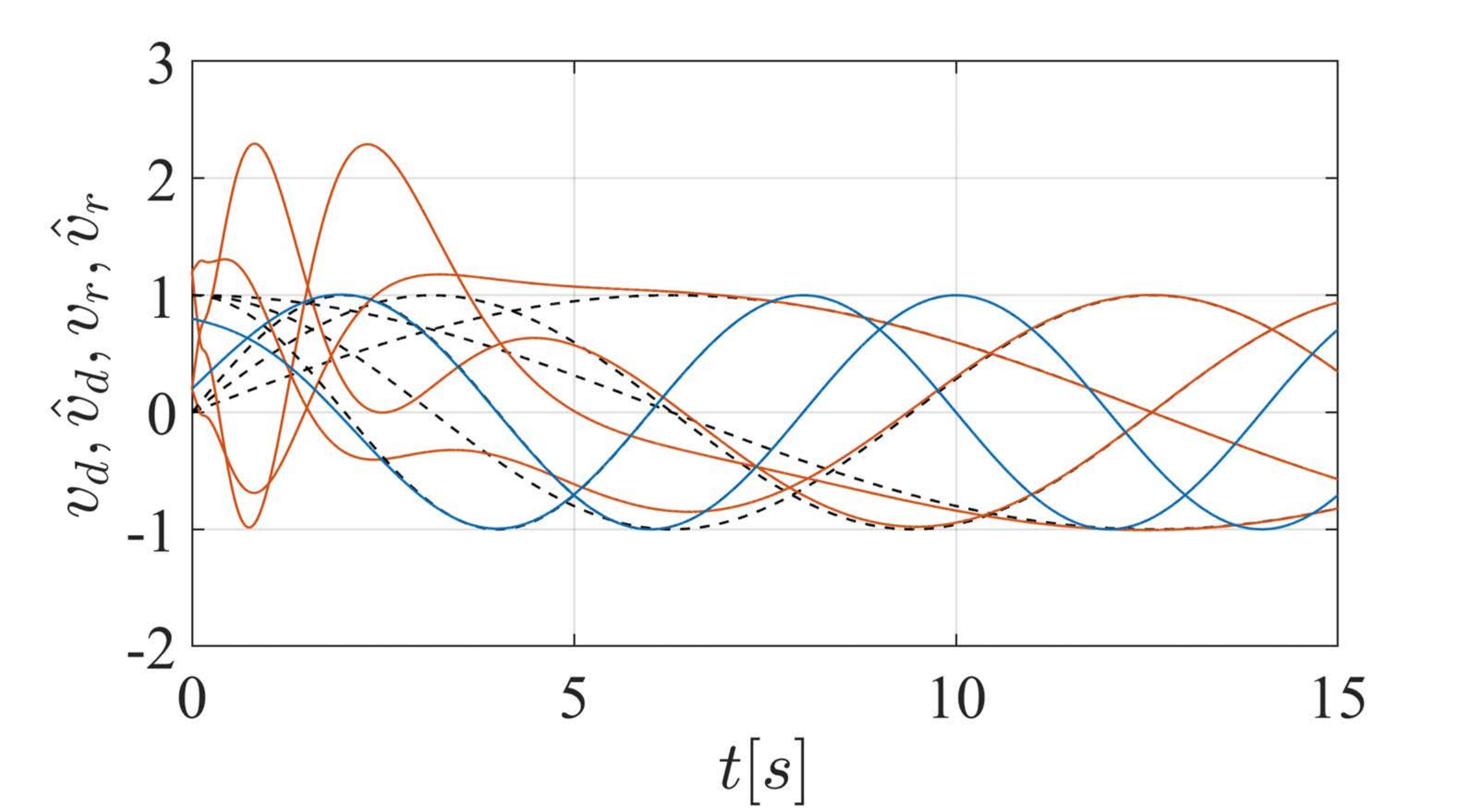}}
  \centerline{(a) Estimates of $\hat v_d$ and $\hat v_r$}
\end{minipage}
\hfill
\begin{minipage}{.49\linewidth}
  \centerline{\includegraphics[width=4.4cm]{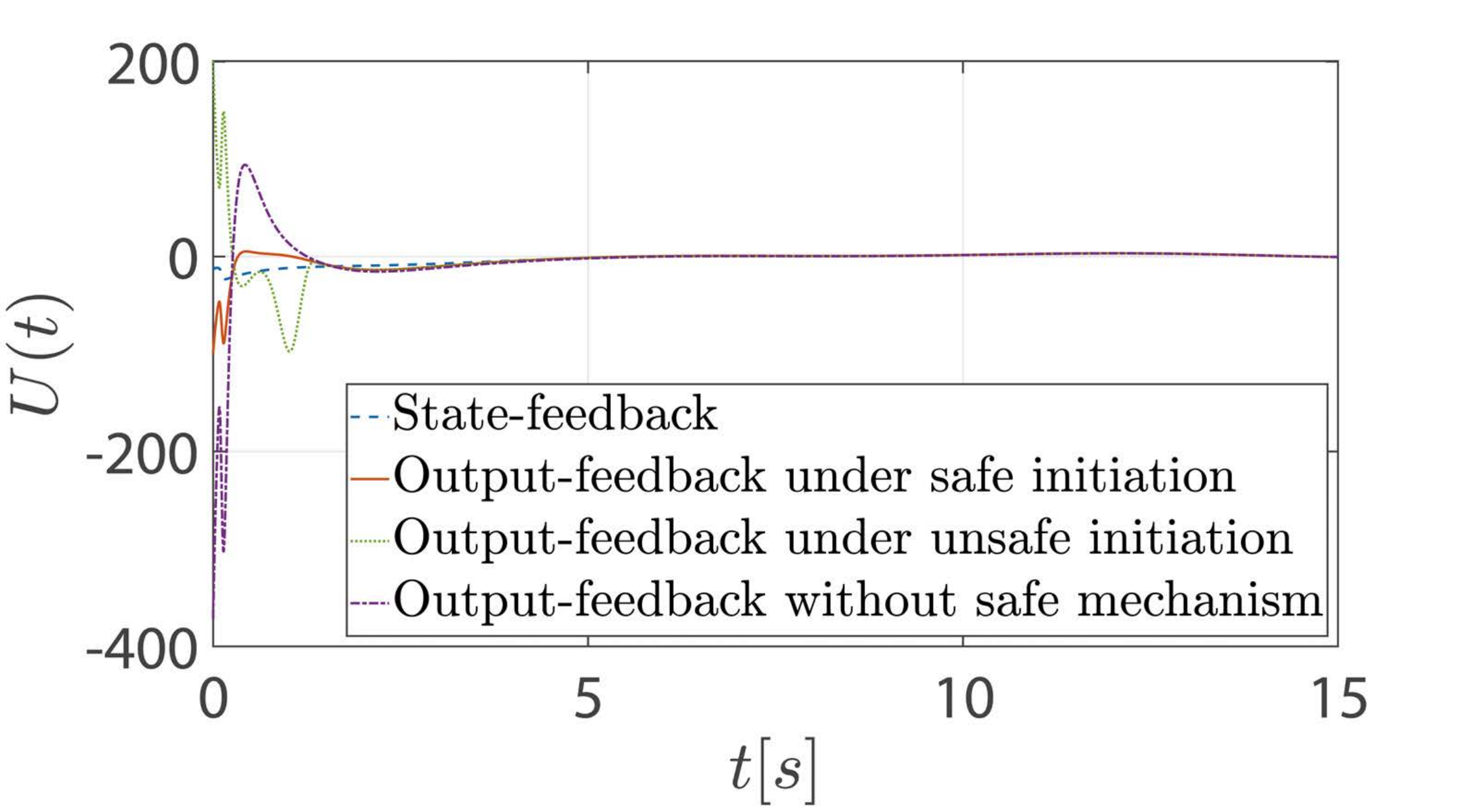}}
  \centerline{(b) Control inputs}
\end{minipage}
\caption{Estimates of exogenous signals $v_d$, $v_r$ given in Sec. \ref{sec:simmod} (red and blue lines are estimates $\hat v_d$ and $\hat v_r$, respectively,  and dashed black lines are their true values) as well as the control inputs.}
\label{fig:es}
\end{figure}

The simulation results are presented below. We know from Fig. \ref{fig:y1} that the plant is open-loop unstable, while both state-feedback control and output-feedback control achieve effective safe regulation: tracking the reference and ensuring the position of the payload stays within the safe region. Moreover, when $y_1$ starts outside the safe region, the designed control input effectively guides it back to safety within 0.77 seconds, which is less than the designated time $\bar t _0+t_a=1.56$ in the control design. Besides, we know from Fig. \ref{fig:y1} that observer state $\hat y_1$ converges to the true value, i.e., the red line that denotes the evolution of $y_1$, in the same case. As can also be seen from Fig. \ref{fig:y1}, the output-regulation controller without a safety mechanism achieves effective trajectory tracking but violates the safety constraint, where the minimum safety margin is $\min h=-2.5<0$, and the first violation occurs at $t=0.41$ s. In contrast, the minimum margin under the proposed output-feedback safe controller is $\min h=0.0197> 0$. Another ODE state $y_2$, i.e., the payload velocity, is shown in Fig. \ref{fig:y2}. It is also shown in Fig. \ref{fig:y2} that the observer state $\hat y_2$ converges to its true value $y_2$, denoted by the red line, in the same case. The PDE states that reflect the oscillations along the cable are depicted in Fig. \ref{fig:pdestates} (where only the safe-initialization output-feedback case is shown to avoid repetition), demonstrating the boundedness of these signals. The observer errors for PDE states and the exogenous signals can be seen in Figs. \ref{fig:ob}, \ref{fig:es}(a), which show that the observer can successfully track the unmeasured PDE states and the exogenous signals $v_d$ and $v_r$. The control inputs in the above four situations are given in Fig. \ref{fig:es}(b).
\section{Conclusion}
This paper presents a safe output regulation strategy for a class of $2 \times 2$ hyperbolic PDE-ODE dynamics subject to fully distributed disturbances. The resulting controller guarantees that the system output remains within a safe region if the initial condition is safe; otherwise, it is driven back into the safe region within a prescribed time set by the user. All closed-loop signals remain bounded, and the tracking error converges to zero. The effectiveness of the proposed framework is demonstrated through a UAV delivery scenario with a cable-suspended payload, achieving both precise trajectory tracking and reliable collision avoidance under wind disturbances. 
The present observer-based safe output regulation framework is developed based on an accurately known plant model and does not address robustness with respect to plant parameter uncertainties. Such robustness may be achieved through regulator designs based on the Internal Model Principle. However, this extension is nontrivial because the transient error of the internal-model state with respect to its desired steady-state trajectory is coupled with the control input, making it difficult to explicitly compensate for this error in the safe control design. This will be investigated in future work.
\section*{Appendix}
\appendix
\renewcommand{\thesubsection}{\Alph{subsection}}
\subsection{The calculation details in the first transformation}\label{sec:trans1Ap}
\setcounter{equation}{0}
\renewcommand{\theequation}{A.\arabic{equation}}
\textbf{Step. 1}
Considering the first equation in the equation set \eqref{eq:Tran-z} with \eqref{eq:Tz}, \eqref{eq:Tv}, applying \eqref{eq:r}, we directly have $z_1(t)=y_1(t)-r(t)=e(t)$, i.e., \eqref{eq:z1e}.\\
\textbf{Step. 2}
Taking the time derivative of the first equation in the equation set \eqref{eq:Tran-z} with \eqref{eq:Tz}, \eqref{eq:Tv}, applying $Y$-ODE \eqref{eq:o1a} and $v$-ODE \eqref{eq:v}, recalling the second equation in the equation set \eqref{eq:Tran-z} with \eqref{eq:vqrrho11}, \eqref{eq:lam0}, \eqref{eq:lami}  for $i=1$, one thus obtains
$\dot z_1(t)=\dot y_1(t)-P_r Sv(t)=a_{11}y_1(t)+y_2+\acute g_1 v(t)-P_r Sv(t)=\varrho _{1,1}y_1+y_2-(\lambda_1+P_r S)v(t)=z_2(t)$.\\
\textbf{Step. 3}
Taking the time derivative of the second equation in the equation set \eqref{eq:Tran-z} with \eqref{eq:Tz}, \eqref{eq:Tv}, applying $Y$-ODE \eqref{eq:o1a} and $v$-ODE \eqref{eq:v}, recalling the third equation in the equation set \eqref{eq:Tran-z} with \eqref{eq:vqrrho21} and \eqref{eq:lam0}, \eqref{eq:lami} for $i=2$, one obtains
$\dot z_2(t)=\dot y_2(t)+\varrho_{11}\dot y_1(t)-(P_r S+\lambda_1)Sv(t)=a_{2,1}y_1+a_{2,2}y_2+y_3+\acute g_2 v(t)+ \varrho_{11}a_{11}y_1(t)+\varrho_{11}y_2+\varrho_{11}\acute g_1 v(t)-(P_r S+\lambda_1)S v(t)=\varrho _{2,1}y_1+\varrho _{2,2}y_2+y_3-(\lambda_2+P_r S^2)v(t)=z_3(t)$.\\
\textbf{Step. 4}
Similarly, taking the time derivative of the third equation in the equation set \eqref{eq:Tran-z} with \eqref{eq:Tz}, \eqref{eq:Tv}, applying $Y$-ODE \eqref{eq:o1a} and $v$-ODE \eqref{eq:v}, recalling the fourth equation in the equation set \eqref{eq:Tran-z} with \eqref{eq:varrhoi1}--\eqref{eq:varrho11}, \eqref{eq:lam0}, \eqref{eq:lami} for $i=3$, one obtains
$\dot z_3(t)=\dot y_3(t)+\sum_{j=1}^{2}\varrho_{2,j}\dot y_j-(P_r S^2+\lambda_2)Sv(t)
=y_4+(a_{3,1}+\sum_{j=1}^{2}\varrho_{2,j}a_{j,1})y_1+(a_{3,2}+\varrho_{2,1}+\varrho_{2,2}a_{2,2})y_2+(a_{3,3}+\varrho_{2,2})y_{3}+(\acute g_{3}+\sum_{j=1}^{2}\varrho_{2,j}\acute g_j-\lambda_2S) v(t)-P_r S^3v(t)
=y_4+\varrho_{3,1}y_1+\varrho_{3,2}y_2+\varrho_{3,3}y_3-(\lambda_3+P_r S^3)v(t)=z_4(t)$.\\
\textbf{Step. 5}
We make an induction hypothesis: for the transformation \eqref{eq:Tran-z} with \eqref{eq:Tz}, \eqref{eq:Tv}, we have
$\dot z_{i}(t)=z_{i+1}(t)$
for $i\ge 3$, under the coefficients $\varrho_{i,\imath},\imath=1,\cdots,i$ and $\lambda_i$ defined by \eqref{eq:varrhoi1}--\eqref{eq:varrho11}, \eqref{eq:lami}, respectively.
We then prove that the induction step holds.
Taking the time derivative of the $i+1$ th equation in the equation set \eqref{eq:Tran-z} with \eqref{eq:Tz}, \eqref{eq:Tv}, applying $Y$-ODE \eqref{eq:o1a} and $v$-ODE \eqref{eq:v}, we have
\begin{align}
&\dot z_{i+1}(t)=\dot y_{i+1}+\sum_{j=1}^{i}\varrho_{i,j}\dot y_j-(\lambda_{i}+P_rS^{(i)})Sv(t)\notag\\
&=y_{i+2}+\sum_{j=1}^{{i+1}}a_{{i+1},j}y_j+\sum_{j=2}^{{i+1}}\varrho_{i,j-1}y_{j}+\sum_{j=1}^{i}\sum_{\jmath=1}^{j}\varrho_{i,j}a_{j,\jmath}y_\jmath\notag\\
&+[\acute g_{{i+1}}+\sum_{j=1}^{i}\varrho_{i,j}\acute g_j -\lambda_{i}S]v(t)-P_rS^{(i)}Sv(t)\notag\\
&=y_{i+2}+[a_{{i+1},1}+\sum_{j=1}^{i}\varrho_{i,j}a_{j,1}]y_{1}\notag\\&+\sum_{\imath=2}^{i}[a_{{i+1},\imath}+\varrho_{i,\imath-1}+\sum_{j=\imath}^{i}\varrho_{i,j}a_{j,\imath}]y_{\imath}+(a_{{i+1},{i+1}}+\varrho_{i,i})y_{{i+1}}\notag\\&+[\acute g_{{i+1}}+\sum_{j=1}^{i}\varrho_{i,j}\acute g_j -\lambda_{i}S-P_rS^{(i+1)}]v(t).\label{eq:ind}
\end{align}
Recalling $z_{i+2}(t)$ that is defined by the $i+2$ th equation in the equation set \eqref{eq:Tran-z}, for $\dot z_{i+1}(t)=z_{i+2}(t)$ to hold, we obtain that the $\varrho_{i+1,\imath},\imath=1,\cdots,i+1$ and $\lambda_{i+1}$ satisfy \eqref{eq:varrhoi1}--\eqref{eq:varrho11}, \eqref{eq:lami} for $i+1$.
Therefore, the induction step holds.
Recalling the Steps 1--4, it follows that $\dot z_{i}(t)=z_{i+1}(t)$ hold for $i=1,\cdots,n-1$ via the transformation \eqref{eq:Tran-z} with \eqref{eq:Tz}--\eqref{eq:lami}.\\
\textbf{Step. 6}
Taking the time derivative of the $n$ th equation in the equation set \eqref{eq:Tran-z} with \eqref{eq:Tz}, \eqref{eq:Tv}, one obtains
$\dot z_n(t)=\dot y_n+\sum_{j=1}^{n-1}\varrho_{n-1,j}\dot y_j-(\lambda_{n-1}+P_rS^{(n-1)})Sv(t)=\sum_{j=1}^{n} a_{n,j} y_{j}(t)+\acute g_n v(t)+ b w(0,t)+\sum_{j=1}^{n-1}\varrho_{n-1,j}(y_{j+1}+\sum_{\jmath=1}^{j}a_{j,\jmath}y_\jmath+\acute g_j v(t))-(\lambda_{n-1}+P_rS^{(n-1)})Sv(t)$.
We thus arrive at $\dot z_n(t)=b w(0,t)+bK^TY(t)-(\lambda_n+P_rS^n)v(t)$, recalling \eqref{eq:varrhoi1}--\eqref{eq:varrho11}, \eqref{eq:lami} for $n$, which is the $n$ th equation of the equation set \eqref{eq:ZA} considering \eqref{eq:Az}, \eqref{eq:K}, and the definition of $B$ and $\bar G_0$.
\subsection{The calculation of the prediction $Z(t+\frac{x}{q_2}),Y(t+\frac{x}{q_2})$}\label{sec:Yfuture}
\setcounter{equation}{0}
\renewcommand{\theequation}{B.\arabic{equation}}
For \eqref{eq:o1}--\eqref{eq:o4}, applying \eqref{eq:Tran-z}, and
\begin{align}
&\eta (x,t) =w(x,t) -\int_0^x {\Psi}(x,y)z(y,t)dy \notag\\
&-\int_0^x {\Phi}(x,y)w(y,t)dy-\lambda(x)Y(t)-\bar\lambda(x)v(t)\label{eq:eta}
\end{align}
where the kernels ${\Psi}$, ${\Phi}$, $\lambda$, $\bar\lambda$ are the same as ones that will be given in Appendix \ref{sec:ker}, i.e., \eqref{eq:traker1}, \eqref{eq:traker2} with \eqref{eq:F}--\eqref{eq:lamx}, and \eqref{eq:ker3}, \eqref{eq:barlam1}, \eqref{eq:barlam2},
we then have
\begin{align}
&\dot Z(t) = A_zZ(t) + B\eta (0,t),\label{eq:eta1}\\
&{\eta_t}(x,t) = q_2{\eta_x}(x,t),\label{eq:eta2}\\
&{\eta (1,t)} {=w(1,t) -\int_0^1 {\Psi}(1,y)z(y,t)dy}\notag\\
& -\int_0^1 {\Phi}(1,y)w(y,t)dy-\lambda(1)Y(t)-\bar\lambda(1)v(t).
\end{align}
We thus have
$Z(t+a)=e^{A_za}Z(t)+\int_0^{a} e^{A_z(a-\tau)}B\eta (q_2\tau,t) d\tau$
for $0\le a\le \frac{1}{q_2}$, i.e.,
\begin{align}
Z(t+a)=&e^{A_za}Z(t)+\frac{1}{q_2}\int_0^{aq_2} e^{A_z(a-\frac{\ell}{q_2})}B\bigg(w(\ell,t) \notag\\&- \int_0^{\ell} {\Psi}(\ell,y)z(y,t)dy-  \int_0^{\ell} {\Phi}(\ell,y)w(y,t)dy\notag\\&-\lambda(\ell)Y(t)-\bar\lambda(\ell)v(t)\bigg)d\ell\notag\\:=&\wp_1(Z(t),z[t],w[t],v(t),a),\label{eq:Zfuture}
\end{align}
for $0\le a\le \frac{1}{q_2}$.
Recalling \eqref{eq:v}, \eqref{eq:Tran-z}, we have
\begin{align}
Y(t+a)=&{T_z}^{-1}Z(t+a)-{T_z}^{-1}T_vv(t+a)\notag\\
=&{T_z}^{-1}\wp_1(T_zY(t)+T_vv(t),z[t],w[t],v(t),a)\notag\\&-{T_z}^{-1}T_ve^{Sa}v(t)\notag\\:=&\wp(Y(t),z[t],w[t],v(t),a),~0\le a\le \frac{1}{q_2},\label{eq:Yfuture}
\end{align}
recalling \eqref{eq:eta2}, \eqref{eq:eta}. Therefore,  $Y(t+\frac{x}{q_2})$, $x\in[0,1]$ can be determined by $Y(t)$, $z[t]$, $w[t]$, $v(t)$ as \eqref{eq:Yfuture} with replacing $a$ by $\frac{x}{q_2}$. \subsection{Kernel conditions of $\varphi,\phi,\Psi,\Phi,\gamma,\lambda,\bar\gamma,\bar\lambda$ in \eqref{eq:contran1a}, \eqref{eq:contran1b}}\label{sec:ker}
\setcounter{equation}{0}
\renewcommand{\theequation}{C.\arabic{equation}}
\subsubsection{Kernel conditions}
By  mapping  \eqref{eq:ZA}, \eqref{eq:o2}--\eqref{eq:o4} and \eqref{eq:targ5}--\eqref{eq:targ4} with using \eqref{eq:o1}, the conditions of the kernels $\varphi,\phi,\Psi,\Phi,\gamma$ and $\lambda$ in the backstepping transformation \eqref{eq:contran1a}, \eqref{eq:contran1b} are obtained as the following equations
\begin{align}
&{q_2}{{\varphi }_y}(x,y) - {q_1}{{\varphi }_x}(x,y) - {d_1}\phi (x,y)=0,\label{1}\\
&{  {q_1}{{\phi }_x}(x,y) {+} {q_1}{{\phi }_y}(x,y)} {+} {d_2}\varphi (x,y)=0,\\
& {q_2}{{\Psi }_x}(x,y)  - {q_1}{{\Psi }_y}(x,y)  - {d_2}\Phi (x,y)=0,\label{eq:kerf0}\\
&{{q_2}{\Phi _x}(x,y)+{q_2}{\Phi _y}(x,y)  - {d_1}\Psi (x,y)}=0,\label{eq:kerf1}
\end{align}
evolving in the triangular domain $\{(x,y):0\le y\le x\le 1\}$ with the boundary conditions
\begin{align}
&\varphi (x,x)=\frac{d_1}{{q_1} + {q_2}},~\Psi (x,x)=\frac{-d_2 }{{q_1} + {q_2}} ,\label{eq:ker1}\\
& \phi (x,0)= \frac{1}{{q_1}p}({q_2}\varphi (x,0)-{\gamma}(x)B) ,\\
& \Phi (x,0) =\frac{1}{{q_2}} (\lambda (x)B+ {q_1}p\Psi (x,0)),\label{eq:Phi0}
\end{align}
where ${\gamma}(x)$, $\lambda (x)$ satisfy
\begin{align}
&  {q_1}{\gamma}'(x) +{\gamma}(x)A + {q_1}{C}\phi (x,0)=0,\label{eq:kerf}\\
& {q_2}\lambda '(x)-{\lambda}(x)A - {q_1}{C}\Psi (x,0)=0,  \\
&{\lambda (0)=-K^T},\label{eq:ker3}\\
& {{\gamma}(0)=-pK^T+C},\label{eq:ker6}
\end{align}
with $K^T$ defined in \eqref{eq:K}.
The equation set \eqref{1}--\eqref{eq:ker6} is a well-known system of heterodirectional linear coupled hyperbolic PDEs-ODE system. Please see \cite{Meglio2017Stabilization} for its well-posedness.

The functions $\bar\lambda$ and $\bar\gamma$ satisfy the following two ODEs
\begin{align}
&\bar\lambda (0)=\frac{1}{b}(\lambda_n+P_rS^{n}),\label{eq:barlam1}\\
&q_2\bar\lambda' (x)-  \bar\lambda (x)S-\lambda(x)\acute G_1+\acute G_3(x)-  \int_0^x {\Phi}(x,y)\acute G_3(y)dy\notag\\& - \int_0^x {\Psi}(x,y)\acute G_2(y)dy-q_1{\Psi}(x,0)\acute G_4=0,\label{eq:barlam2}\\
&\bar \gamma(0)= \frac{p}{b}(\lambda_n+P_rS^{n})-\acute G_4,\\
&-q_1\bar \gamma'(x) -\bar \gamma(x)S-\gamma(x)\acute G_1+\acute G_2(x)- \int_0^x {\phi}(x,y)\acute G_2(y)dy\notag\\& -  \int_0^x {\varphi}(x,y)\acute G_3(y)dy-q_1{\phi}(x,0)\acute G_4=0.\label{eq:kerz}\end{align} The explicit solutions are given below. 
\subsubsection{Explicit solutions}
Equations \eqref{eq:barlam1}--\eqref{eq:kerz} constitute two independent linear
matrix initial-value problems with respect to the spatial variable
$x$. By the variation-of-constants formula, their unique solutions are ($q_1q_2b\neq0$):
\begin{align}
&\bar\lambda(x)
=\frac{1}{b}\left(\lambda_n+P_rS^n\right)
e^{\frac{x}{q_2}S}\notag\\
&+\frac{1}{q_2}\int_0^x
\Bigg[
\lambda(\xi)\acute G_1-\acute G_3(\xi)
+\int_0^\xi\Phi(\xi,y)\acute G_3(y)dy \notag\\
&
+\int_0^\xi\Psi(\xi,y)\acute G_2(y)\,dy
+q_1\Psi(\xi,0)\acute G_4
\Bigg]
e^{\frac{x-\xi}{q_2}S}d\xi,
\label{eq:barlambdasol}\\
&\bar\gamma(x)
=\left[
\frac{p}{b}\left(\lambda_n+P_rS^n\right)-\acute G_4
\right]e^{-\frac{x}{q_1}S}\notag\\
&-\frac{1}{q_1}\int_0^x
\Bigg[
\gamma(\xi)\acute G_1-\acute G_2(\xi)
+\int_0^\xi\phi(\xi,y)\acute G_2(y)\,dy \notag\\
&
+\int_0^\xi\varphi(\xi,y)\acute G_3(y)dy
+q_1\phi(\xi,0)\acute G_4
\Bigg]
e^{-\frac{x-\xi}{q_1}S}d\xi.
\label{eq:bargammasol}
\end{align}The control law requires the solution of $\Psi(x,y)$, $\Phi(x,y)$, whose explicit solution is obtained as follows
according to \cite{supp} where \cite{Vazquez2014Marcum} has been used:
\begin{align}
&\Psi(x,y)={F}(x,y)+\int_y^x L(x,r){F}(r,y)dr,\label{eq:traker1}\\
&\Phi(x,y)={H}(x,y)-L(x,y)+\int_y^x L(x,r){H}(r,y)dr,\label{eq:traker2}
\end{align}
where
\begin{align}
&F(x,y)=\frac{-1}{p(q_1+q_2)}\bigg[\frac{d_1q_2}{pq_1}I_0\left(\frac{2\sqrt{d_1d_2}}{q_1+q_2}\sqrt{(x-y)(\frac{q_1}{q_2}x+y)}\right)
\notag\\&+\sqrt{d_1d_2\frac{x-y}{\frac{q_1}{q_2}x+y}}I_1\left(\frac{2\sqrt{d_1d_2}}{q_1+q_2}\sqrt{(x-y)(\frac{q_1}{q_2}x+y)}\right)
\notag\\&+(pd_2-\frac{d_1q_2}{pq_1})\Pi\left(\frac{pq_1d_2}{q_2}\frac{x-y}{q_1+q_2},\frac{d_1}{pq_1}\frac{q_1x+q_2y}{q_1+q_2}\right)\bigg],\label{eq:F}\\
&H(x,y)=\frac{-1}{q_1+q_2}\bigg[\frac{d_1}{p}I_0\left(\frac{2\sqrt{d_1d_2}}{q_1+q_2}\sqrt{(x-y)(\frac{q_1}{q_2}x+y)}\right)
\notag\\&+\sqrt{d_1d_2\frac{\frac{q_1}{q_2}x+y}{x-y}}I_1\left(\frac{2\sqrt{d_1d_2}}{q_1+q_2}\sqrt{(x-y)(\frac{q_1}{q_2}x+y)}\right)
\notag\\&+(\frac{pd_2q_1}{q_2}-\frac{d_1}{p})\Pi\left(\frac{pq_1d_2}{q_2}\frac{x-y}{q_1+q_2},\frac{d_1}{pq_1}\frac{q_1x+q_2y}{q_1+q_2}\right)\bigg]\label{eq:H}
\end{align}
with  $I_j(j \ge 0)$ denoting the modified Bessel function of the first kind (of order
$j$), and $\Pi(s_1,s_2)=e^{s_1+s_2}(1-s_2e^{-s_1}\int_0^1e^{-\tau s_2}I_0(2\sqrt{\tau s_1s_2})d\tau)$,
and where
\begin{align}
L(x,y)=\frac{-1}{q_2}  \lambda(x-y) {B}\label{eq:traker3}
\end{align}
with $\lambda$, whose initial value is given by \eqref{eq:ker3}, satisfying
\begin{align}
&{q_2}\lambda '(x)-{\lambda}(x)A - {q_1}{C}\int_0^x \frac{-1}{q_2}  \lambda(z) {B}{F}(x-z,0)dz\notag\\&- {q_1}{C}{F}(x,0)=0\label{eq:lamx}
\end{align}
which is a linear Volterra integro-differential equation. By applying the variation-of-constants formula to its differential part, it can be converted into a Volterra integral equation of the second kind, whose unique solution can be obtained by successive approximations.


\end{document}